%% file: main.tex
\RequirePackage{pdfmanagement}
\SetKeys[document/metadata]{
    pdfstandard=A-2b,
    pdfversion=1.7,
    lang=en-US
}
\PassOptionsToPackage{table}{xcolor}
\documentclass[sigconf, nonacm]{acmart}

\usepackage{pvldb}

\renewcommand\vldbdoi{XX.XX/XXX.XX}   
\renewcommand\vldbpages{XXX-XXX}      
\renewcommand\vldbvolume{20}
\renewcommand\vldbissue{4}
\renewcommand\vldbyear{2026}
\renewcommand\vldbpagestyle{empty}
\renewcommand\vldbavailabilityurl{https://github.com/Saddlerdroph/Hybrid-PDP}

\usepackage[normalem]{ulem}
\usepackage{booktabs}
\usepackage{xcolor}
\usepackage{tablefootnote}
\usepackage{csquotes}
\usepackage[ruled]{algorithm2e}
\usepackage{multirow}
\usepackage{graphicx}
\usepackage{tabularx}
\usepackage{float}
\usepackage{enumitem}
\usepackage{soul}
\usepackage{amsmath,amsfonts,amsthm}
\usepackage{threeparttable}
\usepackage{subcaption}
\usepackage{hhline}

\DontPrintSemicolon

\definecolor{inkgreen}{RGB}{0,160,0}
\definecolor{mygray}{HTML}{E7E7E7}
\definecolor{deepgreen}{RGB}{0,100,0}

\AtBeginDocument{%
  }

\newcommand{\bbd}[1]{\textcolor{black}{#1}}

\newcommand{\gbd}[1]{\textcolor{black}{#1}}

\newcommand{\xinghe}[1]{\textcolor{black}{#1}}

\newtheorem{theorem}{Theorem}[section]
\newtheorem{lemma}{Lemma}

\newtheorem{definition}{Definition}[section]

\begin{document}

\title{Hybrid-PDP: Simultaneously Enforcing both Record-level and User-level Personalized Differential Privacy}

\author{Hao Ding}
\authornote{These authors contributed equally to this research.}
\affiliation{%
  \institution{Nanyang Technological University}
  \city{Singapore}
  \country{Singapore}
}
\email{hao028@e.ntu.edu.sg}

\author{Xinghe Chen}
\authornotemark[1]
\affiliation{%
  \institution{Nanyang Technological University}
  \city{Singapore}
  \country{Singapore}
}
\email{xinghe001@e.ntu.edu.sg}

\author{Wei Dong}
\authornote{Wei Dong is the corresponding author.}
\affiliation{%
  \institution{Nanyang Technological University}
  \city{Singapore}
  \country{Singapore}
}
\email{wei\_dong@ntu.edu.sg}

\begin{abstract}
Traditional differential privacy (DP) offers two levels of privacy protection across a dataset, namely \textit{record-level DP} and \textit{user-level DP}. Both of these protections are applied uniformly across records or users, failing to accommodate the heterogeneous privacy requirements that arise in practice. To address this limitation, personalized differential privacy (PDP) was introduced to support non-uniform privacy budgets. However, existing work focuses on either record-level PDP or user-level PDP, without enabling simultaneous personalized protection at both levels.
In practice, such hybrid privacy protection is essential under the PDP setting: assigning personalized privacy budgets disrupts the traditional dominance of user-level protection over individual records, particularly in realistic scenarios where users contribute multiple records and datasets may be aggregated from multiple parties.
Motivated by these limitations, we introduce \textit{hybrid personalized differential privacy} (Hybrid-PDP), a new privacy model that simultaneously enforces personalized protections at both the record- and user-levels. We develop a general framework for achieving Hybrid-PDP, built upon two novel primitives.
The first, \textit{prefix-truncation}, breaks the bidirectional privacy dependencies between records and users.
The second, \textit{doubly-truncation}, builds on prefix-truncation to address unbounded privacy loss arising from heterogeneous record-level budgets.
Furthermore, doubly-truncation mitigates the stringent constraints imposed by user-level personalization.
Extensive experimental results demonstrate the high utility of our framework.
For count, sum, and max queries, our framework achieves an average error reduction of $16.56\times$ and a median reduction of $6.61\times$ compared with the best-performing baseline in each setting.
\end{abstract}

\maketitle

\vldbtopmatter

\input{main_text/1_introduction}
\input{main_text/2_related_work}
\input{main_text/3_preliminary}
\input{main_text/4_Hybrid-PDP}

\input{main_text/5_methodology}
\input{main_text/7_experiment}

\input{main_text/8_conclusion}


\clearpage
\bibliographystyle{ACM-Reference-Format}
\bibliography{reference}

\clearpage
\input{appendix}

\end{document}

%% file: main_text/1_introduction.tex
\section{Introduction}\label{sec:intro}

\textit{Differential Privacy} (DP)~\cite{dwork2006calibrating} provides a rigorous mathematical framework to protect individuals' information in private data releases.
It has been widely adopted across the industry~\cite{apple_dp_2017, ding2017collecting, erlingsson2014rappor} and national statistical agencies~\cite{machanavajjhala2008privacy}.
Roughly speaking, DP ensures that the output distribution remains $\varepsilon$-indistinguishable whether any single individual’s data is included or excluded, where $\varepsilon$ is called the privacy budget and controls the level of protection: a smaller $\varepsilon$ provides a stronger privacy guarantee.
Standard DP adopts two different policies: \textit{record-DP}~\cite{dwork2014algorithmic}, which protects individual data records; \textit{user-DP}~\cite{amin2019bounding,ghazi2023user,dwork2010differentialnew,dong2023continual}, which protects all of a user's data records as a whole.
Consider a banking scenario where a dataset consists of financial records of a bank's customers, such as their transactions, deposits, and credit card repayments.
Record-level DP protects each individual financial record, whereas user-level DP protects all information corresponding to a single customer.
It is trivial to see that, with the same privacy budget, user-DP provides stronger privacy protection.

\begin{figure}[t]
    \centering
    \includegraphics[width=\columnwidth]{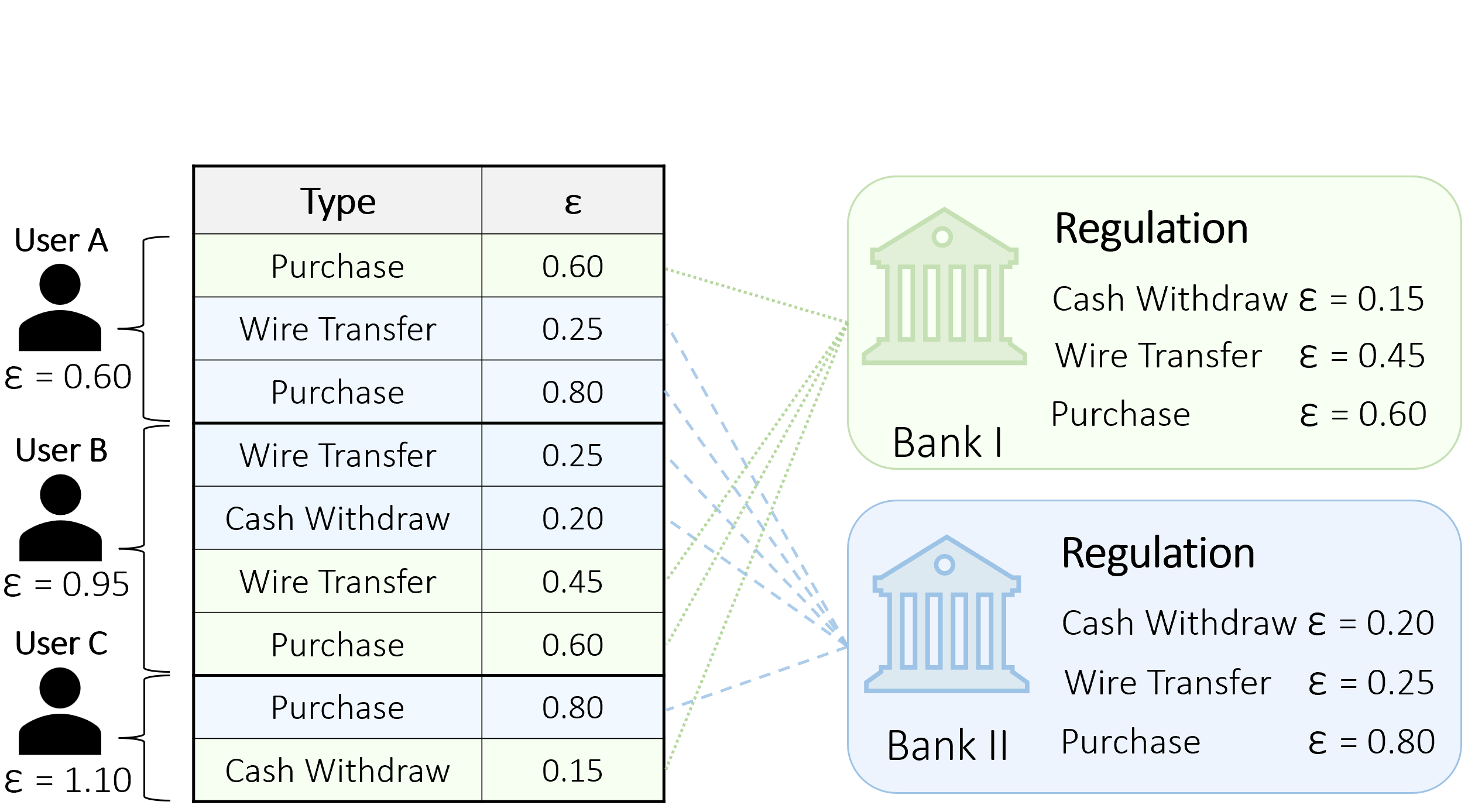}
    \caption{A bank example under Hybrid-PDP.}
    \label{fig:bankexample}
\end{figure}

Although such standard DP policies are widely adopted, the assumption of uniform privacy protection for all records or users may not be desirable.  
Following the banking example, it is common that different customers have their own privacy budgets due to policy considerations associated with their occupations, and different categories of financial records also naturally fall under different privacy regulations or internal policies across different banks.  
For example, a casual coffee purchase transaction may require less protection than a large cash deposit.
To accommodate such diverse privacy requirements, \textit{personalized DP} (PDP) was proposed~\cite{jorgensen2015conservative,alaggan2015heterogeneous,ebadi2015differential}, which allows personalized privacy budgets across a dataset.  
Similar to the standard DP model, current PDP research also adopts two levels of privacy protections: \textit{record-PDP}~\cite{chen2025general, jorgensen2015conservative, liu2024cross} and \textit{user-PDP}~\cite{dajun2024PDP}. 
Under these models, each record or user can have its own privacy budget to better reflect real-world needs.

However, a limitation of these two PDP models is that neither can simultaneously provide personalized protection at both the record- and user-levels. 
Unlike in the standard DP setting, where user-level DP strictly dominates record-level DP, there is no such dominance relationship between user-PDP and record-PDP.
Intuitively, this is because when both records and users have their own personalized privacy budgets, and each user can own multiple records, we cannot guarantee that user-level privacy protection always dominates the record-level one.
This is especially true when the data are contributed by multiple parties, each with their own privacy policies.  
For example, as shown in Figure \ref{fig:bankexample}, when a dataset contains financial records from multiple banks, each bank may have different privacy regulations for different categories of financial records.  
In this case, a single bank cannot determine a customer's personalized privacy budget to sufficiently protect all of that customer's financial data across all banks.

Motivated by these limitations, we introduce \textit{hybrid personalized DP} (Hybrid-PDP), which provides personalized privacy protection at both the user and record levels simultaneously.
Let $\mathcal{U}$ and $\mathcal{R}$ denote the domains of users and records, respectively.
We define a record-level privacy budget function $\Phi_{\mathrm{rec}}: \mathcal{R} \to [\check{\varepsilon}_{\mathrm{rec}}, \hat{\varepsilon}_{\mathrm{rec}}]$ and a user-level privacy budget function $\Phi_{\mathrm{user}}: \mathcal{U} \to [\check{\varepsilon}_{\mathrm{user}}, \hat{\varepsilon}_{\mathrm{user}}]$, where $[\check{\varepsilon}_{\mathrm{rec}}, \hat{\varepsilon}_{\mathrm{rec}}]$ and $[\check{\varepsilon}_{\mathrm{user}}, \hat{\varepsilon}_{\mathrm{user}}]$ denote the predefined ranges of record- and user-level privacy budgets, respectively.
A mechanism satisfies $(\Phi_{\mathrm{rec}}, \Phi_{\mathrm{user}})$-Hybrid-PDP if its output distribution is simultaneously $\Phi_{\mathrm{rec}}(r)$-indistinguishable under the addition or removal of any record $r \in \mathcal{R}$, and $\Phi_{\mathrm{user}}(u)$-indistinguishable under the addition or removal of any subset of records belonging to any user $u \in \mathcal{U}$, including all records of that user.
Notably, under this definition, both privacy budget functions and the domains of users and records are public information, whereas the presence or absence of any particular user or record in the dataset is hidden and protected.
This is consistent with existing notions of user-PDP~\cite{dajun2024PDP} and record-PDP~\cite{chen2025general,jorgensen2015conservative}, while jointly accounting for both levels of protection.

One straightforward approach here is to use a uniform privacy budget at the record-level, which can be derived from some global bounds.  
More specifically, we use the smaller of $\check{\varepsilon}_{\mathrm{rec}}$ and $\check{\varepsilon}_{\mathrm{user}}/\hat{m}$, where $\hat{m}$ is the largest number of records that can be owned by a user.  
However, this approach inevitably introduces an unacceptable error that is tied to global bounds, which are extreme values that typically do not occur in most datasets.
A more desirable error should be independent of such global bounds. Instead, we would like to have an ``instance-specific'' error, which depends on $\varepsilon_{\min}(D)\! =\! \min\{ \varepsilon_{\min}^{\mathrm{rec}}(D),\min_{u \in D }\frac{\Phi_{\mathrm{user}}(u)}{m(D,u)}\}$, where $ \varepsilon_{\min}^{\mathrm{rec}}(D)$ and $\min_{u\in D}\frac{\Phi_{\mathrm{user}}(u)}{m(D,u)}$ can be regarded as instance-specific versions of $\check{\varepsilon}_{\mathrm{rec}}$ and $\check{\varepsilon}_{\mathrm{user}}/\hat{m}$ (see Section~\ref{sec:notation} and ~\ref{sec:strawman} for the formal definitions of $m(D,u)$ and $\varepsilon_{\min}(D)$).
Indeed, reducing the error dependency from global parameters to instance-specific ones is also a key goal of prior record-PDP~\cite{chen2025general} and user-PDP~\cite{dajun2024PDP} works.

However, achieving such instance-specific error bounds is highly challenging.
A natural idea is to estimate the above parameters under both record-PDP and user-PDP.
Yet this approach inevitably violates either record-level or user-level privacy requirements, due to the ``bidirectional'' privacy dependency between records and users.
More precisely, running a record-PDP mechanism to estimate certain record-level parameters also incurs a corresponding privacy loss for the user who owns those records.
Symmetrically, running a user-PDP mechanism to estimate user-level parameters inevitably induces additional privacy loss for the individual records associated with that user.
In addition to the above challenges, a solution for this new hybrid privacy model should be general to cater to all query types.

This ultimately leads to the main research problem of this work:
\begin{quote}
    \textit{``Does there exist a general Hybrid-PDP solution that can be used to answer arbitrary queries while achieving an instance-specific error bound?''}
\end{quote}

\subsection{Our Results}\label{sec:results}
\label{sec:our-results}
\xinghe{In this paper, we give a positive answer to the above research problem by presenting a novel and general framework that transforms any existing record-DP mechanism into a mechanism satisfying Hybrid-PDP.
The key challenge in achieving this transformation is the bidirectional privacy dependency between records and their owning users, which cannot be handled by directly combining existing record-PDP and user-PDP mechanisms.
To address this challenge, we first develop \textit{prefix-truncation}, in which we delete records for each user until the cumulative privacy budget of the retained records does not exceed that user's privacy budget. 
Here, we prioritize deleting the records with larger privacy budgets.}
This ensures that the insertion or deletion of a single record in the initial dataset affects the truncated result by at most two records.
After this process, record-level privacy dominates the user-level privacy.
\gbd{However, under personalized privacy budgets, these two records may have very different budgets, potentially resulting in unbounded privacy loss in subsequent steps.
Consequently, bounding the record distance between neighboring datasets alone is insufficient to bound privacy loss and ensure Hybrid-PDP.
In addition, \textit{prefix-truncation} may remove too many records when user-level privacy constraints are relatively more stringent than those at the record-level.}
\gbd{To address these issues, we further introduce \textit{doubly-truncation}.
Given a threshold $\tau$, we first remove records with privacy budgets below $\tau$ and upgrade the privacy protection of the remaining records by reducing their budgets to $\tau$.
We then apply \textit{prefix-truncation}, thereby bounding the privacy loss induced by additional record changes during this step.}
After the doubly-truncation, all retained records have a privacy budget of $\tau$, and each user's privacy constraint is dominated by that of the records corresponding to the user.
Thus, we can invoke a record-DP mechanism with this budget to satisfy Hybrid-PDP.
This resolves the utility issue of prefix-truncation by choosing a good threshold $\tau$.
Finally, we complete this story by further proposing a Laplace-based threshold selection mechanism with a carefully calibrated score function to find a good $\tau$.  
With this idea, we present a general framework such that, for any query $Q$, given a record-DP mechanism, we construct a corresponding Hybrid-PDP solution whose error depends on $\varepsilon_{\min}(D)$.

In summary, our contributions are threefold:
\begin{itemize}
    \item \xinghe{To the best of our knowledge, this is the first DP notion that simultaneously enforces record-level and user-level personalized privacy constraints, distinguishing it from prior PDP work that considers personalization at only one of these two levels, better addressing real-world privacy needs.}

    \item \xinghe{
    As our main result, we present a general framework for transforming an arbitrary existing record-DP mechanism into a mechanism for our Hybrid-PDP model while providing an instance-specific error guarantee. Specifically, given a record-DP mechanism $\mathcal{M}$ with error 
    $\mathrm{Err}_{\mathcal{M}}\!(D,\varepsilon_{\min}(D))$ on $D$,
    our framework achieves an error bounded by   $\tilde{O}\!(\mathrm{Err}_{\mathcal{M}}\!(D,$ $\varepsilon_{\min}(D)))$
    \footnote{$\tilde{O}(\cdot)$ denotes asymptotic notation that suppresses polylogarithmic factors.}
    and runs in time $\tilde{O}(M + N)$. Here, $M$ and $N$ denote the numbers of users and records in $D$, respectively.}

    \item \bbd{We conduct a comprehensive empirical study of the practical performance of our proposed framework across real-world and synthetic datasets.
    The results demonstrate that our framework achieves an average reduction of \gbd{$16.56\times$} and a median reduction of \gbd{$6.61\times$} in error across all evaluated settings compared with the best-performing baseline in each setting, while maintaining high efficiency.}
    
\end{itemize}

%% file: main_text/2_related_work.tex
\section{Related Work}\label{sec:related}
\xinghe{The motivation for providing privacy protection at different granularities is well established in the traditional non-personalized DP literature, where existing studies typically focus on either record-DP~\cite{dwork2006calibrating,dwork2009differential,dwork2014algorithmic,nissim2007smooth,qardaji2013understanding} or user-DP~\cite{amin2019bounding,epasto2020smoothly,levy2021learning,ghazi2023user}.
Some studies consider both notions, but they generally develop two separate mechanisms, one for each privacy granularity~\cite{liu2020learning, ghazi2023user,dwork2010differential}.
To the best of our knowledge, Pufferfish privacy~\cite{kifer2014pufferfish} and Blowfish privacy~\cite{he2014blowfish} are the two existing privacy notions that can provide privacy guarantees at heterogeneous granularities.
In particular, user-level and record-level neighboring relationships can be defined by customizing the secret set or privacy policies of these notions.
Notably, prior work has not explored enforcing privacy protection simultaneously at both levels because, in the traditional DP setting, a single privacy budget is assigned to the entire mechanism.
Under the same privacy budget, user-DP strictly dominates record-DP, leaving no way to express distinct protection requirements at the two granularities.}

\xinghe{
Turning to the PDP setting, the strength of privacy protection can vary across records and users~\cite{jorgensen2015conservative,ebadi2015differential,alaggan2015heterogeneous,li2017partitioning,niu2020utility,niu2021adapdp,dajun2024PDP,chen2025general,zhang2019probabilistic,meng2019towards,boenisch2023have,lu2024reinforcement,liu2024cross,yuan2025personalized}.
What distinguishes PDP from the traditional DP setting is that the ``user-dominates-record'' argument no longer holds.
Consequently, existing studies explore either record-level PDP~\cite{jorgensen2015conservative,ebadi2015differential,li2017partitioning,niu2020utility,niu2021adapdp,liu2024cross,boenisch2023have} or user-level PDP~\cite{dajun2024PDP}, leaving a gap in providing the desired real-world personalized privacy protection simultaneously at both levels.
Indeed, even in federated learning, where users or clients and records arise naturally, existing PDP studies consider only one level of personalization: either user-level PDP~\cite{lange2025federated,xu2026optimal,liu2023echo,yang2021federated} or record-level PDP~\cite{liu2024cross,liu2022privacy}.
This motivates our work, which, to the best of our knowledge, is the first to address the under-studied hybrid PDP problem.
}

%% file: main_text/3_preliminary.tex
\section{Preliminary}
\subsection{Notation}\label{sec:notation}
Let $\mathcal{U}$ denote the possibly infinite domain of users. For each user $u \in \mathcal{U}$, let $\mathcal{R}_u$ denote the possibly infinite domain of records associated with $u$. The global record domain is then
$
\mathcal{R} = \bigcup_{u \in \mathcal{U}} \mathcal{R}_u .
$
Given a dataset $D$, for any user $u \in \mathcal{U}$, let $D(u) \subseteq \mathcal{R}_u$ denote the set of records contributed by $u$ to $D$. Let
$
m(D,u) = |D(u)|
$
denote the number of records owned by user $u$ in $D$. Then,
$
D = \bigcup_{u \in \mathcal{U}} D(u).
$
Moreover, we write $u \in D$ if user $u$ contributes at least one record to $D$, i.e., $m(D,u)>0$.

For two datasets $D$ and $D'$, the \textit{record-level distance} $d_{\mathrm{rec}}(D,D')$ is the minimum number of record insertions or deletions required to transform $D$ into $D'$.
We say $D$ and $D'$ are $r$-neighbors (denoted $D \sim_r D'$) if $d_\mathrm{rec}(D, D')=1$ and they differ by record $r\in \mathcal{R}$.
Similarly, the \textit{user-level distance} $d_{\mathrm{user}}(D,D')$ is the minimum number of user-level insertions or deletions required to transform $D$ into $D'$, where each user-level operation inserts or deletes all or any subset of records associated with a single user.
We say $D$ and $D'$ are $u$-neighbors (denoted $D \sim_u D'$) if $d_\mathrm{user}(D,D')=1$, meaning the two datasets differ only in the records associated with a certain user $u$.
Furthermore, for clarity of notation, we use $\widetilde{val}$ to denote the privacy-preserving estimation of any true value $val$.

\xinghe{
Besides, some papers~\cite{fang2022shifted,dajun2024PDP,fang2024privacy} consider a more advanced ownership model between users and records, in which each record can be jointly owned by multiple users. 
In this paper, we focus on the basic ownership model, in which each record is owned by only one user, and leave joint ownership for future work.}

\subsection{Differential Privacy}
\begin{definition}[Differential privacy]\label{def:dp}
For any $\varepsilon > 0$, a randomized mechanism $\mathcal{M}$ with output space $\mathcal{Y}$ satisfies $\varepsilon$-differential privacy (DP) if, for any pair of neighboring datasets $D \sim D'$ and any possible output set $Y\subseteq \mathcal{Y}$ of $\mathcal{M}$, 
\begin{equation*}\Pr\bigl[\mathcal{M}(D) \in Y\bigr]
\le
e^{\varepsilon} \Pr\bigl[\mathcal{M}(D') \in Y\bigr].
\end{equation*}
\end{definition}

By integrating user-neighbors and record-neighbors, we can achieve record-level or user-level privacy protection, referred to as \textit{record-DP} and \textit{user-DP}, respectively.
DP satisfies several desirable properties, as outlined below.
\begin{lemma}[Group privacy]\label{lem:group}
If a mechanism $\mathcal{M}$ with output space $\mathcal{Y}$ satisfies $\varepsilon$-DP, then for any two datasets $D$ and $D'$ with $d(D,D')=\lambda$ (i.e., differing by $\lambda$ records or users), and any possible output set $Y\subseteq \mathcal{Y}$ of $\mathcal{M}$, the following holds:
\begin{equation*}
\Pr\bigl[\mathcal{M}(D) \in Y\bigr]
\le
e^{\lambda\varepsilon} \Pr\bigl[\mathcal{M}(D') \in Y\bigr].
\end{equation*}
\end{lemma}

\begin{lemma}[Post-Processing]\label{lem:postprocessing}
If a mechanism $\mathcal{M}$ is $\varepsilon$-DP, then for any mechanism $\mathcal{M}'$, the composed output $\mathcal{M}'(\mathcal{M}(D))$ is still $\varepsilon$-DP.
\end{lemma}

\begin{lemma}[Sequential Composition \cite{dwork2014algorithmic}]\label{lem:sequential}
Let $\mathcal{M}_1$ and $\mathcal{M}_2$ be two randomized mechanisms, where $\mathcal{M}_1$ satisfies $\varepsilon_1$-DP and $\mathcal{M}_2$ satisfies $\varepsilon_2$-DP. Then, $\mathcal{M}(D)=\mathcal{M}_2(D,\mathcal{M}_1(D))$
preserves $(\varepsilon_1+\varepsilon_2)$-DP.
\end{lemma}

A commonly used DP mechanism is the \textit{Laplace mechanism}.  
Given any query $Q : \mathcal{D} \to \mathbb{R}$, its global sensitivity $\mathrm{GS}_Q$ is defined as
$
  \mathrm{GS}_Q = \max_{D \sim D'} \left| Q(D) - Q(D') \right|.
$

\begin{lemma} [Laplace Mechanism \cite{dwork2014algorithmic}]\label{lem:laplace}
Given a query $Q : \mathcal{D} \to \mathbb{R}$, the Laplace mechanism
$\mathcal{M}(D) \;=\; Q(D) + \eta$
satisfies $\varepsilon$-DP, where $\eta$ is drawn from a Laplace distribution with scale
$\mathrm{GS}_Q/\varepsilon$.
\end{lemma} 

Note that all the aforementioned properties and mechanisms are applicable to both record-DP and user-DP by adopting the corresponding notations under a specific DP policy. 
For example, under the record-DP setting, we employ the record-level distance $d_{\mathrm{rec}}(D, D')$ in Lemma~\ref{lem:group} and the record-neighboring relation in Lemma~\ref{lem:postprocessing} and Lemma~\ref{lem:sequential}.

\subsection{Personalized Differential Privacy}\label{sec:pdp}
Standard DP assigns a uniform privacy budget to all units, e.g., records or users. 
However, this assumption is often unrealistic. 
For example, in a banking scenario, confidential transfers and routine transactions may require different levels of privacy protection.

To accommodate such heterogeneous privacy requirements, prior work proposed \textit{personalized differential privacy} (PDP)~\cite{jorgensen2015conservative}, which assigns different privacy budgets to different records or users. 
Under PDP, we have a privacy function $\Phi$ to assign the privacy budget for a unit $x$, which can be either a user or a record.

\begin{definition}[Personalized Differential Privacy (PDP)]
For any privacy function $\Phi$, a randomized mechanism $\mathcal{M}$ with output space $\mathcal{Y}$ satisfies $\Phi$-PDP if, for any pair of neighboring datasets $D \sim_x D'$ and any possible output set $Y\subseteq \mathcal{Y}$ of $\mathcal{M}$, the following holds:
\begin{equation*}
\Pr\bigl[\mathcal{M}(D)\in Y\bigr]
\le
e^{\Phi(x)} \Pr\bigl[\mathcal{M}(D') \in Y\bigr].
\end{equation*}
\end{definition}

Similar to standard DP, PDP also supports both record-level and user-level privacy protection. 
\gbd{Moreover, the group privacy property extends naturally to PDP:}
\gbd{\begin{lemma}[PDP Group privacy]\label{lem:pdpgroup}
If a mechanism $\mathcal{M}$ with output space $\mathcal{Y}$ satisfies $\Phi$-PDP, then for any two datasets $D$ and $D'$ with $D \sim_{\mathcal{S}} D'$, where $\mathcal{S}=\{x_1,x_2,\ldots,x_n\}$ is the set of differing individuals, and any possible output set $Y\subseteq \mathcal{Y}$ of $\mathcal{M}$, the following holds:
\begin{equation*}
\Pr\bigl[\mathcal{M}(D) \in Y\bigr]
\le
e^{\sum_{x\in\mathcal{S}}\Phi(x)}
\Pr\bigl[\mathcal{M}(D') \in Y\bigr].
\end{equation*}
\end{lemma}}

By letting $x$ denote either a user $u$ or a record $r$, we obtain \textit{user-PDP} and \textit{record-PDP}, respectively.

\paragraph{Limitations of PDP}Although PDP supports heterogeneous privacy, it can be applied only at a single level, e.g., either user-level or record-level. However, in practice, both users and records may simultaneously require personalized protection.
In the banking example, each customer may have different privacy requirements due to occupation-related policies: some occupations require stricter confidentiality, while others allow higher disclosure risk.
Meanwhile, different transactions may exhibit varying levels of privacy sensitivity, influenced by factors such as the merchant category, associated banking policies, or time of occurrence, which are independent of customer-level policies.
These observations motivate the introduction of a novel PDP model that satisfies two-level heterogeneous privacy requirements.

Indeed, this privacy issue, requiring user- and record-level privacy guarantees simultaneously, is unique to the PDP setting and does not arise in standard DP. In standard DP, user-level DP always dominates record-level DP: an $\varepsilon$-user-DP mechanism naturally provides at least $\varepsilon$-record-DP, since adding or removing a user also covers the addition or removal of any individual record owned by that user.
However, PDP introduces additional challenges. Records may have different privacy budgets, and the user-record ownership relation, that is, whether a record associated with a given user exists in the dataset, is itself private. As a result, user-PDP does not necessarily dominate record-PDP.



%% file: main_text/4_Hybrid-PDP.tex
\section{Hybrid Personalized Differential Privacy}

Motivated by the limitations of existing PDP frameworks, we propose a novel privacy model called \textit{Hybrid Personalized Differential Privacy} (Hybrid-PDP), which provides heterogeneous privacy protection at both the record- and user-levels. 
The detailed mathematical formulation of Hybrid-PDP and its desired properties are presented in Section~\ref{sec:hdp-defapro}. 
In Section~\ref{sec:strawman}, we introduce a straightforward idea to achieve Hybrid-PDP for arbitrary queries. 
However, such an idea can offer only limited utility, as we elaborate later.

\subsection{Definition and Basic Properties}\label{sec:definition}
\label{sec:hdp-defapro}

To provide both user- and record-level privacy protection simultaneously, we define two public privacy functions, $\Phi_{\mathrm{rec}}: \mathcal{R} \to [\check{\varepsilon}_{\mathrm{rec}},$ $ \hat{\varepsilon}_{\mathrm{rec}}]$ and $\Phi_{\mathrm{user}}: \mathcal{U} \to [\check{\varepsilon}_{\mathrm{user}}, \hat{\varepsilon}_{\mathrm{user}}]$, respectively, where $[\check\varepsilon_\mathrm{rec}, \hat\varepsilon_\mathrm{rec}]$ and $[\check\varepsilon_\mathrm{user}, \hat\varepsilon_\mathrm{user}]$ are predefined ranges
with conservatively small lower bounds and large upper bounds, respectively. 
More precisely, each user $u$ is assigned a privacy budget of $\Phi_\mathrm{user}(u)$, and each record $r$ is assigned a privacy budget of $\Phi_\mathrm{rec}(r)$.
For later use, we also define
$
\hat\varepsilon=\min(\hat{\varepsilon}_\mathrm{rec}, \hat{\varepsilon}_\mathrm{user})$ and $\check\varepsilon =\min(\check{\varepsilon}_\mathrm{rec}, \check{\varepsilon}_\mathrm{user} / \hat{m})
$, where $\hat{m}$ denotes a public finite contribution bound such that each user can contribute at most $\hat{m}$ records to any dataset. Our Hybrid-PDP model can be formalized as:

\begin{definition}[Hybrid Personalized Differential Privacy]\label{def:hpdp}
A randomized mechanism $\mathcal{M}$ with output space $\mathcal{Y}$ satisfies $(\Phi_\mathrm{rec},$ $\Phi_\mathrm{user})$-Hybrid-PDP if both of the following conditions hold for any possible output set $Y\subseteq \mathcal{Y}$ of $\mathcal{M}$:

\begin{itemize}
  \item \textbf{Record-level constraint.} For every record $r \in \mathcal{R}$ and every pair of $r$-neighboring datasets $D \sim_{r} D'$,
  \begin{equation*}
    \Pr\bigl[\mathcal{M}(D) \in Y\bigr]
    \le
    e^{\Phi_\mathrm{rec}(r)} \,
    \Pr\bigl[\mathcal{M}(D') \in Y\bigr].
  \end{equation*}

  \item \textbf{User-level constraint.} For every user $u \in \mathcal{U}$ and every pair of $u$-neighboring datasets $D \sim_{u} D'$,
  \begin{equation*}
    \Pr\bigl[\mathcal{M}(D) \in Y\bigr]
    \le
    e^{\Phi_\mathrm{user}(u)} \,
    \Pr\bigl[\mathcal{M}(D') \in Y\bigr].
  \end{equation*}
\end{itemize}
\end{definition}

\paragraph{Remark.}
\xinghe{Notably, the privacy functions as well as the domains of users and records are public.
However, whether a particular user or record is present in a dataset $D$ remains private.
Hence, the budgets associated with the users and records actually present in $D$ are not revealed.
This setting is consistent with the privacy model adopted in existing PDP studies~\cite{jorgensen2015conservative,dajun2024PDP,chen2025general}.
Moreover, when privacy budgets are correlated with sensitive information, e.g., record values or user attributes, the data curator can create multiple public copies of the same record, each associated with a different possible privacy budget. The private dataset contains only one of these copies, while which copy is included remains private. This implementation therefore keeps both the record's actual privacy budget and any sensitive information correlated with it private.}
\gbd{We formalize this deployment, including the information observable to external analysts, 
in Appendix~\ref{app:private-budgets}.}

\begin{lemma}[Hybrid-PDP Post-processing]\label{lem:hdp-post}
If a mechanism $\mathcal{M}$ is $(\Phi_\mathrm{rec},\Phi_\mathrm{user})$-Hybrid-PDP, then for any mechanism $\mathcal{M}'$, the composed output $\mathcal{M}'(\mathcal{M}(D))$ is still $(\Phi_\mathrm{rec},\Phi_\mathrm{user})$-Hybrid-PDP.
\end{lemma}

\begin{lemma}[Hybrid-PDP Sequential Composition]\label{lem:hdp-comp}
Let $\mathcal{M}_1$ and $\mathcal{M}_2$ be two randomized mechanisms such that $\mathcal{M}_1$ satisfies $(\Phi_{\mathrm{rec}}^1, \Phi_{\mathrm{user}}^1)$-Hybrid-PDP and $\mathcal{M}_2$ satisfies $(\Phi_{\mathrm{rec}}^2, \Phi_{\mathrm{user}}^2)$-Hybrid-PDP. Then,
$\mathcal{M}(D)=\mathcal{M}_2(D,\mathcal{M}_1(D))$
satisfies $(\Phi_{\mathrm{rec}}, \Phi_{\mathrm{user}})$-Hybrid-PDP, where for all $r \in \mathcal{R}$ and $u \in \mathcal{U}$,
$
\Phi_\mathrm{rec}(r) = \Phi_\mathrm{rec}^1(r) + \Phi_\mathrm{rec}^2(r), 
$ $
\Phi_\mathrm{user}(u) = \Phi_\mathrm{user}^1(u) + \Phi_\mathrm{user}^2(u).
$
\end{lemma}

The proofs of the two lemmas above are similar to the corresponding proofs for standard DP~\cite{dwork2014algorithmic}.
Additionally, to facilitate later analysis, we introduce the following instance-wise extremal quantities under the Hybrid-PDP framework.
For a dataset $D$ with user- and record-level privacy functions $(\Phi_{\mathrm{user}}, \Phi_{\mathrm{rec}})$, define
$\varepsilon_{\min}^{\mathrm{user}}(D) := \min_{u \in D} \Phi_\mathrm{user}(u)$, $m_{\max}(D) := \max_{u \in D} m(D,u)$ and $\varepsilon_{\min}^{\mathrm{rec}}(D) := \min_{r \in D} \Phi_\mathrm{rec}(r).$
Notably, since the presence or absence of records and users is private, all the above parameters are private and should be protected.

\subsection{Straw-man: Utilizing Global Bounds of Privacy Parameters to Achieve Hybrid-PDP}\label{sec:strawman}

Here, we propose a straightforward baseline mechanism for achieving $(\Phi_\mathrm{rec}, \Phi_\mathrm{user})$-Hybrid-PDP. 
The high-level idea is to reduce the Hybrid-PDP problem to a record-DP problem, enabling the use of existing record-level mechanisms to answer the query. 
The key challenge lies in deriving a uniform record-level privacy budget that simultaneously ensures the two-level protection guarantee of Hybrid-PDP.

Naively, such a uniform privacy budget can be derived from the predefined global parameters $\check{\varepsilon}_{\mathrm{rec}},\; \check{\varepsilon}_{\mathrm{user}}$ and $\hat{m}$.
Specifically, we use a uniform global record-level budget $\check{\varepsilon} = \min\!\left(\check{\varepsilon}_{\mathrm{rec}},\; \check{\varepsilon}_{\mathrm{user}} / \hat{m}\right)$. It is straightforward to show that this approach naturally satisfies both the record-level and user-level privacy requirements defined by the $(\Phi_\mathrm{rec}, \Phi_\mathrm{user})$-Hybrid-PDP guarantee.

(i) \textit{Record-level privacy guarantee.} For any $r \in \mathcal{R}$, we have $\check{\varepsilon} \le \Phi_\mathrm{rec}(r)$.  
Thus, this mechanism is $\check{\varepsilon}$-DP at the record level, and therefore satisfies the record-level privacy requirement of $\Phi_\mathrm{rec}$.

(ii) \textit{User-level privacy guarantee.} For any $u \in \mathcal{U}$, we have $\hat{m} \check{\varepsilon} \le \Phi_\mathrm{user}(u)$. 
Since any user can own at most $\hat{m}$ records, by the group privacy property (Lemma~\ref{lem:group}), the total privacy loss of any user is bounded by $\hat{m} \check{\varepsilon}$. 
Thus, it satisfies the user-level privacy requirement specified by $\Phi_\mathrm{user}$.

Moving toward the utility analysis, recall that for any query $Q$, we reduce the Hybrid-PDP problem to a record-DP problem, and a record-DP mechanism $\mathcal{M}$ is involved.
Here, we first formalize the error of $\mathcal{M}$ as $\mathrm{Err}_{\mathcal{M}}\!\left(D, \varepsilon, \beta\right)$:
for a given dataset $D$, privacy budget $\varepsilon$, with probability at least $1-\beta$, we have
$    |Q(D)-\mathcal{M}(D)| \leq \mathrm{Err}_{\mathcal{M}}\!\left(D, \varepsilon, \beta\right)$.
Then, we have 
\begin{lemma}
For any $\beta \in [0,1)$, privacy budget functions $\Phi_{\mathrm{rec}}$ and $\Phi_{\mathrm{user}}$, and dataset $D$, by applying a record-DP mechanism $\mathcal{M}$ with error $\mathrm{Err}_{\mathcal{M}}(\cdot)$ as defined above and setting $\check{\varepsilon}$ as the privacy budget, the naive baseline achieves, with probability at least $1-\beta$, an error of
$\mathrm{Err}_{\mathcal{M}}\!\left(D,\check{\varepsilon},\beta\right)$.
\end{lemma}

\paragraph{Utility limitation}
As we have shown, $\check{\varepsilon}$ is solely determined by the predefined global parameters rather than the specific data instance. 
This is often undesirable, as the global parameters tend to be overly conservative in order to cover all privacy requirements across the record and user universe. 
It leads to significant utility degradation, as most existing standard DP mechanisms have error inversely proportional to the privacy budget, which in our case becomes $1/\check{\varepsilon}$.

Instead, a more desirable error should have some instance-specific dependency. We would like to achieve an error of
\begin{equation}\label{eq:ins-err}
    \mathrm{Err}_{\mathcal{M}}\!\left(D, c\cdot\varepsilon_{\min}(D), \beta\right), \quad \beta \in[0,1)
\end{equation}
where $c$ is a constant and $\varepsilon_{\min}(D)$ is defined as
\begin{equation}
\label{eq:ins-min}
  \varepsilon_{\min}(D)
  \;:=\;
  \min\!\Bigl\{
     \varepsilon_{\min}^{\mathrm{rec}}(D),\;
     \min_{u \in D }(\Phi_{\mathrm{user}}(u)/m(D,u))
  \Bigr\}.
\end{equation}
Roughly speaking, $\varepsilon_{\min}(D)$ reflects the most stringent actual privacy requirement, whether user- or record-level, present in the queried dataset $D$. 
Thus, it leads to an instance-specific error for $D$.

To provide intuition on how this improves utility, consider the bank example. 
The globally strongest privacy requirements for certain transactions or customers may be extremely small, resulting in $\check{\varepsilon} \sim 10^{-9}$. 
However, the actual value of $\varepsilon_{\min}(D)$ in a specific bank database is often orders of magnitude larger, as such sensitive transactions or customers may not appear in that bank.

%% file: main_text/5_methodology.tex
\section{A Hybrid-PDP framework with instance-specific error}\label{sec:methodology}
In this section, we study how to achieve instance-specific error under Hybrid-PDP.
In Section~\ref{attempt}, we present two straightforward attempts. 
Although these attempts fail to satisfy Hybrid-PDP, they provide the foundation for our final solution. 
The difficulty arises from privacy dependencies between records and users, which prevent us from processing records and users independently.
To eliminate these privacy dependencies, in Section~\ref{sec:prefix} we introduce a \textit{prefix-truncation} mechanism. 
\gbd{However, due to the heterogeneity of record-level privacy requirements, prefix-truncation alone cannot guarantee Hybrid-PDP.
This motivates a more comprehensive doubly-truncation mechanism in Section~\ref{sec:doub-trunc}, which addresses this issue by upgrading the privacy protection of certain records.
}
Next, in Section~\ref{sec:thld-select}, we show how doubly-truncation can serve as a building block for a threshold-selection algorithm under Hybrid-PDP. 
By designing a private scoring function for each candidate threshold, the algorithm selects a threshold that balances the trade-off between truncating too many records and using privacy budgets that are too small, which leads to excessive noise.
Finally, in Section~\ref{sec:framework}, building on the idea of prefix-truncation, we propose our Hybrid-PDP solution that achieves instance-specific errors for arbitrary queries.

\subsection{Straightforward Attempts}
\label{attempt}

\paragraph{Attempt 1: Estimating instance-specific parameters independently.}
Recall that our target is to achieve an instance-specific error in (\ref{eq:ins-err}), which further depends on three instance-specific values, $\varepsilon_{\min}^{\mathrm{user}}(D)$, $\varepsilon_{\min}^{\mathrm{rec}}(D)$ and $m_{\max}(D)$.
At first glance, the problem seems decomposable: we can estimate these parameters separately by applying the PDP private estimator from prior work~\cite{chen2025general}.
More precisely, at the record-level, we first privately estimate $\varepsilon^{\mathrm{rec}}_{\min}(D)$ using the existing record-PDP solution, and then use it as a threshold to remove records whose privacy budgets fall below it.
Then, similarly, at the user-level, we estimate $\varepsilon^{\mathrm{user}}_{\min}(D)$ and remove users (together with all their records) whose privacy budgets fall below the user-level threshold. After this step, since all users have privacy budgets at least $\tilde{\varepsilon}^{\mathrm{user}}_{\min}(D)$, we can further estimate $m_{\max}(D)$ using a record-DP maximum estimator, as in \cite{dong2023universal} and \cite{dwork2014algorithmic}, with the uniform privacy budget $\tilde{\varepsilon}^{\mathrm{user}}_{\min}(D)$.
After that, we cap the number of each user's records at the estimated maximum $\widetilde{m}_{\max}(D)$.
Finally, after these three rounds of truncation, we apply a record-DP mechanism with a uniform privacy budget $\varepsilon'=\min\!\left\{\tilde{\varepsilon}^{\mathrm{rec}}_{\min}(D),\,\frac{\tilde{\varepsilon}^{\mathrm{user}}_{\min}(D)}{\widetilde{m}_{\max}(D)}\right\}$ to all records in the resulting dataset.  
When these parameters are estimated with sufficient accuracy, we nearly attain the error bound in $\mathrm{Err}_{\mathcal{M}}\!\left(D, \Omega(\varepsilon'), \beta\right)$.
The quantity $\varepsilon'$ is still smaller than our target $\varepsilon_{\min}(D)$ because it estimates 
$\varepsilon_{\min}^{\mathrm{rec}}(D)$ 
and 
$\varepsilon_{\min}^{\mathrm{user}}(D)$
separately, whereas 
$\varepsilon_{\min}(D)$
accounts for them jointly. 
However, $\varepsilon'$ is instance-specific and can be considerably larger than the global bound $\check{\varepsilon}$.

From a naive perspective, this approach appears to satisfy Hybrid-PDP, since we estimate both the record- and user-level privacy parameters under privacy constraints separately. 
However, it ignores the privacy dependency between users and records arising from the fact that each user may contribute multiple records.
Consequently, when we estimate $\varepsilon_{\min}^{\mathrm{rec}}(D)$, we consume privacy budget not only at the record-level but also at the user-level, which can be accounted for through group privacy (Lemma~\ref{lem:group}). 
For example, suppose a user $u$ with $\Phi_{\mathrm{user}}(u)=0.2$ has three records, each with privacy budget $0.1$.
Then record-level estimation exhausting the record-level budget causes a total privacy loss of $0.1+0.1+0.1=0.3>0.2$ for $u$, which violates the user-level privacy requirement.

\paragraph{Attempt 2: A record-level parameters capping-based estimation method.}
Motivated by the failure of Attempt~1 --- in which estimating the minimal record-level parameter in (\ref{eq:ins-min}) compromises user-level privacy, we refine the procedure by  swapping the order of estimation and adding a capping step.
More precisely, we first privately estimate the user-relevant parameters $\varepsilon_{\min}^{\mathrm{user}}(D)$
and
$m_{\max}(D)$, and then truncate users and their records according to the rules in Attempt~1. 
Then, we further cap each of the record-level privacy budgets by $\tilde{\varepsilon}_{\min}^{\mathrm{user}}(D)/\widetilde{m}_{\max}(D)$. 
After that, we estimate the minimal record-level privacy budget over the clipped dataset. 
Finally, we run a record-DP mechanism on the resulting dataset with that uniform record-level privacy budget.

This idea successfully bounds user-level privacy leakage during estimation of the minimal record-level privacy budget. Since each user contributes at most $\widetilde{m}_{\max}(D)$ records after truncation and each record consumes at most $\tilde{\varepsilon}_{\min}^{\mathrm{user}}(D)/\widetilde{m}_{\max}(D)$ of privacy budget, the total privacy loss for any retained user is at most $\tilde{\varepsilon}_{\min}^{\mathrm{user}}(D)$, thereby controlling user-level privacy loss during estimation of the minimal record-level privacy budget. 
However, this overlooks the fact that privately estimating user-level parameters such as $\varepsilon_{\min}^{\mathrm{user}}(D)$ and $m_{\max}(D)$ also consumes the record-level privacy budget.
For example, if each user has a single record, then the existence of a record implies the existence of the corresponding user. 
Thus, deleting that record may induce a change in $\varepsilon_{\min}^{\mathrm{user}}(D)$.
Moreover, $m_{\max}(D)$ depends on per-user record counts in the dataset $D$.
In the extreme case where the dataset contains only one user, inserting or deleting a record necessarily changes $m_{\max}(D)$.
Therefore, privately estimating $\varepsilon_{\min}^{\mathrm{user}}(D)$ or $m_{\max}(D)$ must be accounted for within the record-level privacy budget.

\paragraph{Discussion}
These attempts reveal a \textit{bidirectional} privacy dependency between the record- and user-levels. 
Concretely, estimating record-level parameters incurs additional user-level loss via group privacy, while estimating user-level parameters can still consume record-level budget since record-neighboring changes may alter the user set.
Therefore, without additional preprocessing of the datasets, any sequential estimation pipeline that starts at either level tends to under-account for the other and fail to satisfy Hybrid-PDP.


\subsection{Prefix-truncation: Dealing with Bidirectional Privacy Dependency}
\label{sec:prefix}
Motivated by the bidirectional privacy dependency, in this section we propose a per-user truncation-based preprocessing step designed to ensure that record-level privacy implies user-level privacy.
For each user $u$, we truncate the records $D(u)$ until the cumulative record-level privacy budget of the retained records is at most $\Phi_{\mathrm{user}}(u)$.
After this preprocessing, the record-level privacy guarantee dominates the user-level guarantee via group privacy (Lemma~\ref{lem:group}).
Here, we first assume an ordering over records. Then, for each user $u$, we retain only the longest prefix (under this order) whose cumulative record-level privacy budget does not exceed $\Phi_{\mathrm{user}}(u)$.
We refer to this idea as \textit{prefix-truncation}.

To perform prefix-truncation, the most straightforward idea is to assume there is a natural order on records (e.g., lexicographic).
However, performing truncation according to such an ordering can violate the desired privacy guarantees.
The issue is that inserting or deleting a single record for a user can change the positions of many of that user’s other records in the ordering, thereby altering which records are retained.
As a result, a small change can cascade into multiple changes in the truncated dataset, causing record-level neighbors to become far apart after truncation.
For example, consider a user with privacy budget $1$ and $100$ records, where each record has a privacy budget of $0.01$.
After prefix-truncation, all records are retained.
Now we construct another record-neighboring dataset by inserting a record for the user that has privacy budget $1$ and precedes all existing records in the global order.
Then this neighboring dataset retains only one record after truncation, leading to an unbounded record-level distance between the original and neighboring truncated datasets.

\begin{figure*}[t]
    \centering
    \includegraphics[width=0.9\textwidth]{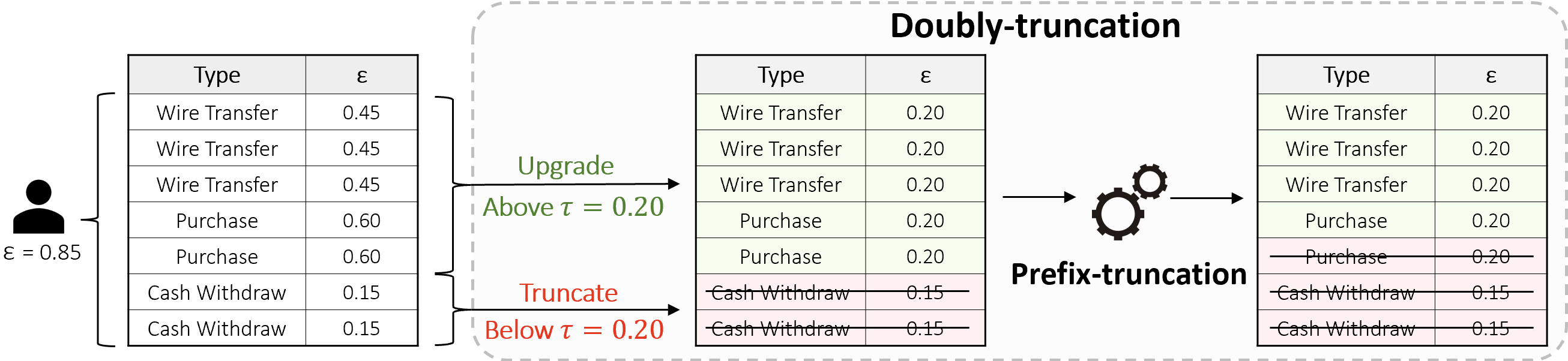}
 \caption{Illustration of the doubly-truncation workflow for a user with $\tau = 0.20$.}
    \label{fig:doubly}
\end{figure*}

To avoid the distance blow-up caused by prefix-truncation under an arbitrary predefined order, we instead introduce a \textit{privacy-based ordering}.
Specifically, we sort the records in $\mathcal{R}$ in nondecreasing order of $\Phi_{\mathrm{rec}}(r)$, breaking ties according to a fixed global natural order.
This ordering is therefore fixed and fully data-independent.
In the following, we show that prefix-truncation under this ordering guarantees that any two record-level neighboring datasets remain within a bounded distance after truncation.
\bbd{Due to space limitations, we include only proof sketches in the main paper and defer the complete proofs of all lemmas and theorems to Appendix~\ref{proof}.} 

\begin{lemma}[Record-level Stability of privacy-based order truncation]\label{lem:privacy-sort}
For any record-neighboring datasets $D\sim_r D'$, let $\overline D$ and $\overline D'$ denote the datasets obtained by applying privacy-based ordering prefix-truncation to $D$ and $D'$, respectively. 
Then the record-level distance $d_{\mathrm{rec}}(\overline D, \overline D') \leq 2$.
\end{lemma}

\begin{proof}[\bbd{Proof Sketch}]
\xinghe{
Intuitively, since prefix-truncation is performed independently for each user, adding or deleting a record affects only its owner's truncation result. We focus on deletion, as addition is symmetric. If the deleted record lies outside the retained prefix, the result is unchanged. Otherwise, its deletion removes one retained record and releases its privacy cost. Since records are ordered by nondecreasing privacy budgets, this released budget can admit at most one additional record into the prefix. Thus, the truncation result differs by at most one deletion and one insertion, implying that $d_{\mathrm{rec}}(\overline{D}, \overline{D}') \le 2$.
}
\end{proof}

In the example discussed earlier, under the privacy-based ordering, the new record with a privacy budget of $1$ cannot precede any of those with a privacy budget of $0.01$.
Therefore, adding it will not cause any of the existing records to be additionally truncated.
\gbd{Moreover, after prefix-truncation, it is straightforward to show that record-level privacy dominates user-level privacy via group privacy (Lemma~\ref{lem:pdpgroup}).}
\begin{lemma}[Dominance of record-level privacy over user-level privacy after prefix-truncation]\label{lem:domin}
Let $\overline D$ denote the dataset obtained by prefix-truncation. Then, for any user $u$, $\sum_{r \in \overline D(u)}\Phi_{\mathrm{rec}}(r) \leq \Phi_{\mathrm{user}}(u)$.
\end{lemma}

\begin{proof}[\bbd{Proof Sketch}]
\xinghe{
The result follows directly from the definition of prefix-truncation. For each user $u$, prefix-truncation retains the longest prefix whose cumulative record-level privacy budget does not exceed $\Phi_{\mathrm{user}}(u)$. Therefore, by construction, $\sum_{r \in \overline D(u)}$ $\Phi_{\mathrm{rec}}(r)$ $ \leq \Phi_{\mathrm{user}}(u)$. If $u$ does not appear in the dataset, then $\overline{D}(u)=\emptyset$, and the inequality holds trivially.
}
\end{proof}

\gbd{Building on the above two properties, one may attempt to achieve Hybrid-PDP by applying \textit{prefix-truncation} followed by a record-PDP mechanism to the retained dataset $\overline{D}$.
However, this approach fails to satisfy Hybrid-PDP due to the heterogeneity of record-level privacy budgets.
Specifically, although Lemma~\ref{lem:privacy-sort} bounds the record-level distance between the retained datasets by $2$, the two differing records may have substantially different privacy budgets.
Consequently, the privacy guarantee derived from group privacy (Lemma~\ref{lem:pdpgroup}) for the downstream record-PDP mechanism may be too weak to satisfy Hybrid-PDP.
To illustrate this issue, consider a dataset containing a user $u$ with $\Phi_{\mathrm{user}}(u)=1$ and a record $r_1$ belonging to $u$ with $\Phi_{\mathrm{rec}}(r_1)=1$, which is retained after \textit{prefix-truncation}.
Now consider a neighboring dataset obtained by inserting another record $r_2$ belonging to $u$, with $\Phi_{\mathrm{rec}}(r_2)=0.01\ll 1$.
This insertion causes \textit{prefix-truncation} to retain $r_2$ instead of $r_1$.
Since the two retained datasets differ in both $r_1$ and $r_2$, applying group privacy (Lemma~\ref{lem:pdpgroup}) to the downstream record-PDP mechanism yields a privacy loss bound of $\Phi_{\mathrm{rec}}(r_1)+\Phi_{\mathrm{rec}}(r_2)=1.01$, which is insufficient to establish the required $0.01$-privacy guarantee.
Moreover, because the two affected records may have arbitrarily different privacy budgets, simply scaling down the downstream privacy budgets by a constant factor (e.g., halving them) cannot generally ensure the required privacy guarantee.
Indeed, in the above example, even halving the budgets fails to provide the required privacy guarantee, since $1.01/2 > 0.01$.
Beyond this privacy issue, \textit{prefix-truncation} may also degrade utility.
When user-level privacy constraints are substantially more stringent than record-level constraints, \textit{prefix-truncation} may discard a large fraction of records, potentially even the entire dataset.
Thus, \textit{prefix-truncation} alone can neither guarantee Hybrid-PDP nor achieve the desired utility.}

\subsection{Doubly-truncation: Adapting Record-level Privacy to User-level Privacy}
\label{sec:doub-trunc}
\gbd{The main issue with prefix-truncation arises from heterogeneous record-level privacy budgets.
As discussed in Section~\ref{sec:prefix}, although privacy-based ordering bounds the record-level distance after truncation, the privacy budgets of the affected records can still differ substantially.
To address this issue, we can assign a uniform privacy budget to all retained records before applying prefix-truncation.
This ensures that any indirectly affected record has the same privacy budget as the originally changed record.
Thus, by group privacy, the total privacy loss is bounded by twice the budget of the downstream record-DP mechanism, allowing us to satisfy the original record-level privacy requirement by halving the budget allocated to the downstream mechanism.
Moreover, assigning a uniform budget can also improve data utility. 
One key observation is that existing general PDP methods~\cite{chen2025general} ultimately reduce the problem to record-DP by applying record-DP techniques under a uniform privacy budget $\tau$ shared by all records. 
Since the privacy loss in the downstream record-DP mechanism is bounded by $\tau$, maintaining larger record-level budgets during prefix-truncation unnecessarily consumes the user's budget without improving downstream utility.
Therefore, we can upgrade the privacy protection of records with $\Phi_{\mathrm{rec}}(r)>\tau$ to a stricter level $\tau$, allowing more records to be retained during prefix-truncation.}


Inspired by the above insights, we propose a \textit{doubly-truncation} strategy (Algorithm~\ref{alg:doub-trun}), in which we first truncate some records under stricter privacy protection, and then upgrade the privacy of the remaining records before applying prefix-truncation. Since all retained records after the first truncation are upgraded to the same privacy budget, using the natural order also satisfies the privacy-based ordering requirement in this case.
We design the following algorithm.
\begin{algorithm}[h]
\LinesNumbered
\caption{
Doubly-truncation}
\label{alg:doub-trun}
\KwIn{Dataset $D$, privacy functions $\Phi_\mathrm{rec}$, $\Phi_\mathrm{user}$, and a threshold $\tau$}
\KwOut{The retained dataset $\overline{\overline D}$}
 \ForEach {$u \in D$}{
  ${D}_{\tau}(u) \gets \left\{r\in D(u) \;:\; \Phi_\mathrm{rec}(r) \ge \tau \right\}$\;
    
  $\kappa(u) \gets \min\!\left( \left\lfloor \frac{ \Phi_\mathrm{user}(u) }{ \tau } \right\rfloor,\; \left|{D}_{\tau}(u)\right| \right)$\;
    
  $\overline{\overline D}(u) \gets$ first $\kappa(u)$ records of ${D}_\tau(u)$ under the natural order\;
}
  $\overline{\overline D} \gets \bigcup_{u \in D} \overline{\overline D}(u)$\;

\Return $\overline{\overline D}$\;
\end{algorithm}

Precisely, for a given threshold $\tau$, we drop all records with privacy budgets below $\tau$, which ensures that all records have privacy budgets at least $\tau$, allowing us to invoke record-DP algorithms with privacy budget $\tau$ to guarantee record-level privacy. 
Then, we upgrade the remaining records' privacy budgets to $\tau$, allowing more records to be retained during the downstream per-user prefix-truncation. 
After that, we perform prefix-truncation for each user to ensure that satisfying the record-level privacy guarantee also implies satisfaction of the user-level privacy guarantee.
Figure~\ref{fig:doubly} illustrates an example of doubly-truncation for a user with $\tau = 0.20$.  
Moreover, doubly-truncation can be shown to satisfy the same two properties as prefix-truncation:

\begin{lemma}[Record-level Stability of doubly-truncation]\label{lem:privacy-sort-dt}
For any record-neighboring datasets $D\sim_r D'$, let $\overline{\overline{D}}$ and $\overline{\overline {D'}}$ denote the datasets obtained by applying doubly-truncation to $D$ and $D'$ with the same truncation threshold. 
Then the record-level distance $d_{\mathrm{rec}}(\overline{\overline {D}}, \overline{\overline {D'}}) \leq 2$.
\end{lemma}

\begin{proof}[\bbd{Proof Sketch}]
\xinghe{
For record-neighboring datasets $D,D'$, the first truncation and privacy upgrade produce datasets that are either identical or record-neighboring. By Lemma~\ref{lem:privacy-sort}, prefix-truncation then yields $d_{\mathrm{rec}}(\overline{\overline D},\overline{\overline{D'}})\leq 2$.
}
\end{proof}

\begin{lemma}[Dominance of record-level privacy over user-lev\\el privacy after doubly-truncation]
\label{lem:domin-dt}
Let $\overline{\overline{D}}$ denote the dataset obtained by applying doubly-truncation with threshold $\tau$. Then, for any user $u$, 
$\sum_{r \in \overline{\overline{D}}(u)} \tau
\leq \Phi_{\mathrm{user}}(u)$.
\end{lemma}

\begin{proof}[\bbd{Proof Sketch}]
\xinghe{
By construction, the final prefix-truncation in doubly-truncation ensures that the total upgraded privacy budget of each user does not exceed its user-level budget. Since each retained record has budget $\tau$, we have $\sum_{r \in \overline{\overline{D}}(u)} \tau \leq \Phi_{\mathrm{user}}(u)$.
}
\end{proof}

\gbd{With these two desired properties, and since all remaining records have been assigned a uniform effective privacy budget $\tau$ before prefix-truncation, we can design a Hybrid-PDP mechanism that first applies doubly-truncation with threshold $\tau$ and then invokes a record-DP mechanism with privacy budget $\tau/2$.
Specifically, the uniform budget allows us to bound the privacy loss arising from at most two changed records (Lemma~\ref{lem:privacy-sort-dt}) via standard group privacy (Lemma~\ref{lem:group}), as both records now have the same privacy budget. Meanwhile, Lemma~\ref{lem:domin-dt} guarantees that the user-level privacy requirement is also satisfied.
}
This approach induces an error trade-off governed by the choice of $\tau$.
Intuitively, an overly large threshold results in more records being dropped, thereby reducing data utility, whereas an overly small threshold introduces excessive DP error in the downstream record-DP mechanisms.
One appropriate choice is to set the threshold to $\varepsilon_{\min}(D)$ as in (\ref{eq:ins-min}). 
With this threshold, all records are retained after the doubly-truncation step, and the DP error introduced by the threshold is controlled by our desired instance-specific quantity. 
Nevertheless, as established in Section~\ref{attempt}, any attempt to explicitly estimate $\varepsilon_{\min}(D)$ inevitably fails due to the bidirectional privacy dependency between the record- and user-levels. 

\subsection{Threshold Selection in Doubly-truncation}
\label{sec:thld-select}
Recall that our goal is to determine a value of $\tau$ for the doubly-truncation process that is as large as possible while still retaining as many records as possible.
To reduce the search space of $\tau$, we consider $\ell=\left\lceil \log\left( \hat{\varepsilon}/\check{\varepsilon} \right) \right\rceil+1
$ candidate thresholds.\footnote{All logarithms denoted by $\log$ are base 2 throughout this paper.} 
More precisely, we set the $i$-th threshold as 
$\tau_i = 2^{i-1}\cdot \check{\varepsilon}$.
For these thresholds, we use a score function to measure their utilities, and we want to privately select a $\tau_i$ with the highest score.

Here, our score function consists of two components: 
the first component is the number of records retained after applying doubly-truncation under the current threshold.
The second component is a penalty term that is inversely proportional to the threshold. 
This term prevents selecting a threshold $\tau$ that is too small, i.e., $\tau \ll \varepsilon_{\min}(D)$ as defined in (\ref{eq:ins-min}), since such a choice would result in adding excessive DP error to the downstream record-DP mechanism.
The formal definition of the score function is given by:
\begin{equation*}
S_i := \left| \overline{\overline{D}}_{i} \right| - \frac{3\ell}{\tau_i} \ln(\ell/\beta)
\end{equation*}
where $\overline{\overline D}_{i}$ denotes the retained dataset after doubly-truncation, while $ -\frac{3\ell}{\tau_i} \ln(\ell/\beta)$ is the penalty term.  
With high probability, as we will show later, this term prevents the privatized score function from selecting an overly small threshold.

We now discuss how to privatize each $S_i$. 
First, it is easy to show that upgrading the privacy and applying prefix-truncation do not enlarge the record-level sensitivity of $S_i$:
\begin{lemma}\label{lem:score}
    For any threshold $\tau_i$, the corresponding score function $S_i$ changes by at most 1 between any pair of record-neighboring datasets $D \sim_r D'$.
\end{lemma}
\begin{proof}[\bbd{Proof Sketch}]
\xinghe{
Lemma~\ref{lem:score} holds because, in the first step of the doubly-truncation procedure, i.e., truncating records with privacy budgets below $\tau_i$,
inserting or deleting any record only affects whether that record is included in the intermediate retained dataset $\overline{D}$.
In the second step, i.e., prefix-truncation, by Lemma~\ref{lem:privacy-sort}, we know that inserting any record can cause zero or one other record to be dropped.
Combining these observations for the two steps, we conclude that, for the entire doubly-truncation procedure, a single insertion or deletion in the original dataset $D$ can change the size of the final retained dataset $\overline{\overline{D}}_i$ by at most $1$.
}
\end{proof}

\begin{algorithm}[t]
\LinesNumbered
\caption{
Threshold Selection}
\label{alg:thld-select}
\KwIn{Dataset $D$, privacy functions $\Phi_\mathrm{rec}$, $\Phi_\mathrm{user}$, and failure rate $\beta$}
\KwOut{Selected threshold $\tau_{i^{\star}}$}
$\ell=\left\lceil \log\left( \hat{\varepsilon}/\check{\varepsilon} \right) \right\rceil+1$\;

\For{$i \gets 1$ \KwTo $\ell$}{
  $\tau_i \gets 2^{i-1} \cdot \check{\varepsilon}$\;
   
$\overline{\overline D}_i \gets \textsc{Doubly-Truncation}\bigl(D, \Phi_\mathrm{rec}, \Phi_\mathrm{user}, \tau_i\bigr)$\;

  $\widetilde{S}_i \gets \left| \overline{\overline D}_{i} \right|\;-\;\frac{3\ell}{\tau_i} \ln(\ell/\beta) + \mathrm{Lap}(\ell/\tau_i) $\;
}

$i^{\star} \gets \arg\max_{1 \le i \le \ell} \;\widetilde{S}_i$\;
\Return $\tau_{i^{\star}}$\;
\end{algorithm}

To privatize all $\ell$ scores, the idea is to add Laplace noise with a scale of $\ell/\tau_i$.
Then, by Lemma~\ref{lem:score}, estimating each score function preserves $\Phi_{\mathrm{rec}}/\ell$-record-PDP. 
Moreover, Lemma~\ref{lem:domin-dt} ensures that any downstream mechanism satisfying record-level privacy also satisfies user-level privacy. 
Therefore, estimating each score satisfies $(\Phi_{\mathrm{rec}}/\ell,\, \Phi_{\mathrm{user}}/\ell)$-Hybrid-PDP.
By sequential composition, the entire estimation process, which has $\ell$ scores, complies with the desired privacy guarantees of $(\Phi_{\mathrm{rec}},\, \Phi_{\mathrm{user}})$-Hybrid-PDP.

\begin{figure}[h!]
    \centering
    \includegraphics[width=0.9\columnwidth]{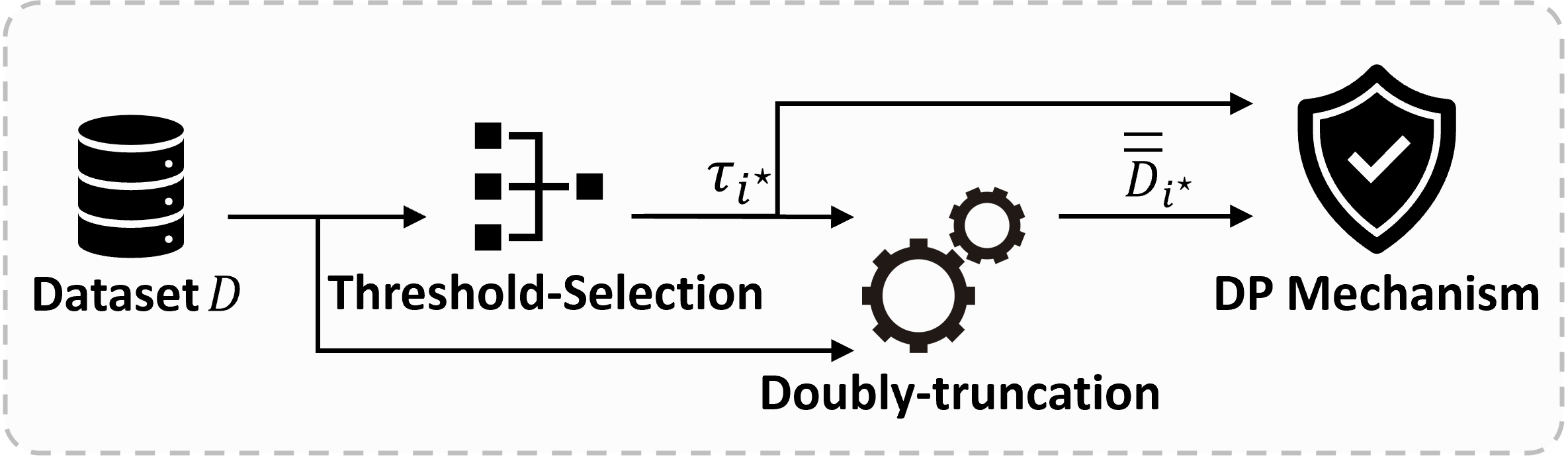}
    \caption{\bbd{Illustration of the workflow of our proposed Hybrid-PDP framework.}}
    \label{fig:workflow}
\end{figure}


To conclude, the privatized score function can be set as
\begin{equation*}
        \widetilde{S}_i :=\left| \overline{\overline{D}}_{i} \right| - \frac{3\ell}{\tau_i} \ln(\ell/\beta) + \mathrm{Lap}(\ell/\tau_i).
\end{equation*}

The detailed algorithm for privately estimating the best threshold is given in Algorithm~\ref{alg:thld-select}.

\begin{theorem}\label{thm:thld-select}
For any given dataset $D$ and any given privacy functions $\Phi_{\mathrm{rec}}$ and $\Phi_{\mathrm{user}}$, Algorithm~\ref{alg:thld-select} satisfies $(\Phi_{\mathrm{rec}},\; \Phi_{\mathrm{user}})$-Hybrid-PDP and, with probability at least $1 - \beta$, outputs $\tau_{i^\star}$ such that the following hold:

(i) $\tau_{i^\star} \ge \tfrac{1}{2}\,\varepsilon_{\min}(D)$.

(ii) $\bigl|\,|D| - |\overline{\overline D}_{i^\star}|\bigr| = O\!\left(\dfrac{1}{\varepsilon_{\min}(D)}\log(\hat{\varepsilon}/\check{\varepsilon}) \log\!\frac{\log(\hat{\varepsilon}/\check{\varepsilon})}{\beta}\right)$, 
where $\overline{\overline D}_{i^\star}$ is obtained by applying doubly-truncation to $D$ with threshold $\tau_{i^\star}$.
\end{theorem}

\begin{proof}[\bbd{Proof Sketch}]
\bbd{
For privacy, consider iteration $i$. If a record $r$ satisfies $\Phi_{\mathrm{rec}}(r)<\tau_i$, it is removed by doubly-truncation and incurs no privacy loss. Otherwise, the score has record-level sensitivity at most one, so Laplace noise with scale $\ell/\tau_i$ yields privacy loss at most $\tau_i/\ell\leq\Phi_{\mathrm{rec}}(r)/\ell$. At the user level, adding or removing a user $u$ changes the score by at most the number of retained records for $u$. By Lemma~\ref{lem:domin-dt}, their total upgraded privacy budget is at most $\Phi_{\mathrm{user}}(u)$, yielding user-level privacy loss at most $\Phi_{\mathrm{user}}(u)/\ell$. Hence, each iteration satisfies $(\Phi_{\mathrm{rec}}/\ell,\Phi_{\mathrm{user}}/\ell)$-Hybrid-PDP, and sequential composition over $\ell$ iterations, together with post-processing, gives the claimed $(\Phi_{\mathrm{rec}},\Phi_{\mathrm{user}})$-Hybrid-PDP guarantee.
}

\bbd{For utility, with probability at least $1-\beta$, all Laplace noises at iteration $i$ are bounded by $(\ell/\tau_i)\ln(\ell/\beta)$. Let $k$ satisfy $\varepsilon_{\min}(D)\in[\tau_k,2\tau_k)$. For all $j\leq k$, doubly-truncation removes no records, so $|\overline{\overline D}_j|=|D|$. Under these noise bounds, the bias term ensures that the privatized score at index $k$ is at least that at every $j<k$, implying $i^\star\geq k$ and hence $\tau_{i^\star}\geq\tau_k>\varepsilon_{\min}(D)/2$. Comparing the selected score with that at index $k$ gives $|D|-|\overline{\overline D}_{i^\star}|\leq(4\ell/\tau_k)\ln(\ell/\beta)\leq(8\ell/\varepsilon_{\min}(D))\ln(\ell/\beta)$, which yields the claimed bound.
}
\end{proof}

In conclusion, using Algorithm~\ref{alg:thld-select}, we can estimate an appropriate threshold that is neither too small, i.e., larger than $\tfrac{1}{2}\,\varepsilon_{\min}(D)$ to ensure that only instance-specific DP noise is introduced in the subsequent record-DP mechanism, nor too large, so that it excludes at most $\tilde{O}\!\left(1/\varepsilon_{\min}(D)\right)$ records from the original dataset.
Thus, we set the selected threshold as a uniform privacy budget over the retained dataset, and then invoke a downstream record-DP mechanism with the selected $\tau_{i^\star}$.

\subsection{Overall Mechanism}\label{sec:framework}

After showing how to select an appropriate $\tau_{i^\star}$ under Hybrid-PDP,
we are now ready to present our final Hybrid-PDP framework.
Given any existing record-DP mechanism, our framework can extend it to Hybrid-PDP.
In this framework, in Stage 1, we run Algorithm~\ref{alg:thld-select} with $\Phi_{\mathrm{rec}}/2$ and $\Phi_{\mathrm{user}}/2$ to obtain a threshold $\tau_{i^\star}$ with the guarantee claimed in Theorem~\ref{thm:thld-select}.
\footnote{
Strictly speaking, Algorithm~\ref{alg:thld-select} first returns 
$\tau^{\mathrm{half}}_{i^\star}$, which satisfies Theorem~\ref{thm:thld-select} 
with respect to the halved privacy functions. 
By setting $\tau_{i^\star}=2\tau^{\mathrm{half}}_{i^\star}$, the guarantees of 
Theorem~\ref{thm:thld-select} translate back to the original privacy functions 
with the same form.}\label{footnote4}
Then, in Stage 2, we perform doubly-truncation with the threshold $\tau_{i^\star}$ over $D$ with Algorithm~\ref{alg:doub-trun} to obtain the truncated dataset $\overline{\overline{D}}_{i^\star}$.
Afterward, we apply the given record-DP mechanism with privacy budget $\tau_{i^\star}/4$.  
The factor \(1/4\) arises from two sources.  First, half of the overall privacy budget is spent in Stage 1 to estimate \(\tau_{i^\star}\). Therefore, in Stage 2, the remaining privacy budget available to each record and each user is reduced by a factor of \(2\), and the effective truncation threshold becomes \(\tau_{i^\star}/2\). Second, doubly-truncation can enlarge the record-level distance by a factor of \(2\). To compensate for this enlargement, the privacy budget of the record-DP mechanism is reduced by another factor of \(2\).
\footnote{\bbd{The distance metric used in Lemma~\ref{lem:privacy-sort-dt} is the insert/delete-one distance. If we consider
the change-one distance, the distance remains \(1\).
Thus, if the given record-DP} \bbd{mechanism works under both the
insert/delete-one and change-one settings, we can use
$\tau_{i^\star}/2$ instead of $\tau_{i^\star}/4$ as the budget for the downstream record-DP mechanism.}}
The detailed algorithm is presented in Algorithm~\ref{alg:Hybrid-PDP}, and Figure~\ref{fig:workflow} illustrates its workflow.

\begin{algorithm}[htpb]
\LinesNumbered
\caption{Instance-Specific Hybrid-PDP Mechanism}
\label{alg:Hybrid-PDP}
\KwIn{A dataset $D$, record-DP mechanism $\mathcal{M}$, privacy budget functions $\Phi_\mathrm{rec}, \Phi_\mathrm{user}$, and failure rate $\beta$}
\KwOut{$\widetilde{Q}(D)$ under $(\Phi_\mathrm{rec}, \Phi_\mathrm{user})$-Hybrid-PDP}

\tcp{Stage 1: Threshold selection:}
$\tau_{i^\star}^{\mathrm{half}} \gets \textsc{ThresholdSelection}(D, \Phi_\mathrm{rec}/2, \Phi_\mathrm{user}/2, \beta/2)$\;
$\tau_{i^\star} \gets 2\tau_{i^\star}^{\mathrm{half}}$\;
\tcp{Stage 2: Run a record-DP mechanism on the truncated dataset:}
$\overline{\overline{D}}_{i^\star}\gets \textsc{Doubly-Truncation}\bigl(D, \Phi_\mathrm{rec}, \Phi_\mathrm{user}, \tau_{i^\star}\bigr)$\;

$\widetilde{Q}(D) \gets \mathcal{M}\bigl(\overline{\overline{D}}_{i^\star}, \tau_{i^\star}/4, \beta/2\bigr)$\;

\Return $\widetilde{Q}(D)$
\end{algorithm}

For efficiency, suppose there are $M$ users and $N$ records.
Algorithm~\ref{alg:thld-select} runs in
$O\!\left(\log\!\left(\hat{\varepsilon}/\check{\varepsilon}\right)(M+N)\right)$
time, because it performs
$(\left\lceil \log\!\left(\hat{\varepsilon}/\check{\varepsilon}\right) \right\rceil+1)$
rounds, each involving a doubly-truncation that takes $O(M+N)$ time.
For privacy, by Theorem~\ref{thm:thld-select}, $\tau_{i^*}$ computed in Stage~1 of
Algorithm~\ref{alg:Hybrid-PDP} satisfies
$(\Phi_{\mathrm{rec}}/2,\Phi_{\mathrm{user}}/2)$-Hybrid-PDP.
Moreover, given $\tau^*$, Lemma~\ref{lem:privacy-sort-dt} and Lemma~\ref{lem:domin-dt}
imply that Stage~2 satisfies
$(\Phi_{\mathrm{rec}}/2,$ $\Phi_{\mathrm{user}}/2)$-Hybrid-PDP.
By sequential composition, Algorithm~\ref{alg:Hybrid-PDP} satisfies
$(\Phi_{\mathrm{rec}},$ $\Phi_{\mathrm{user}})$-Hybrid-PDP.
For utility, we establish the following guarantee:

\begin{theorem}\label{thm:utility}
For any privacy functions $\Phi_\mathrm{user}$ and $\Phi_\mathrm{rec}$, any $\beta \in (0,1)$, and any record-DP mechanism $\mathcal{M}$ for query $Q$, Algorithm~\ref{alg:Hybrid-PDP} satisfies $(\Phi_{\mathrm{rec}}, \Phi_{\mathrm{user}})$-Hybrid-PDP and, with probability at least $1 - \beta$, returns $\widetilde{Q}(D)$ such that the following holds:
\begin{equation*}
\bigl|\widetilde{Q}(D) - Q(D)\bigr|
\;\le\;
\underbrace{\bigl|Q(D) - Q(D')\bigr|}_{I}
\;+\;
\underbrace{\mathrm{Err}_{\mathcal{M}}\bigl(D', \varepsilon_{\min} (D)/8, \beta/2\bigr)}_{II},
\label{eq:utility-bound}
\end{equation*}
where $D'$ is obtained by removing
$
O\!\left(\dfrac{1}{\varepsilon_{\min}(D)}\log(\hat{\varepsilon}/\check{\varepsilon}) \log\!\frac{\log(\hat{\varepsilon}/\check{\varepsilon})}{\beta}\right)
$
records from $D$.
\end{theorem}

\begin{proof}[\bbd{Proof Sketch}]
\xinghe{
Intuitively, applying Algorithm~\ref{alg:thld-select} with the halved privacy functions and failure probability $\beta/2$, Theorem~\ref{thm:thld-select} guarantees, with probability at least $1-\beta/2$, that the threshold satisfies $\tau_{i^\star}\geq\varepsilon_{\min}(D)/2$ and that the retained dataset $D'$ is obtained by removing at most
$O\!\left(\frac{1}{\varepsilon_{\min}(D)}\log(\hat{\varepsilon}/\check{\varepsilon})\log\!\frac{\log(\hat{\varepsilon}/\check{\varepsilon})}{\beta}\right)$
records. 
Moreover, since $\mathcal{M}$ runs on $D'$ with privacy budget $\tau_{i^\star}/4\geq\varepsilon_{\min}(D)/8$, with probability at least $1-\beta/2$, its error is no more than $\mathrm{Err}_{\mathcal{M}}(D',$ $\varepsilon_{\min}(D)/8,$ $\beta/2)$. 
By a union bound, both utility guarantees hold simultaneously with probability at least $1-\beta$. 
The triangle inequality then gives
$|\widetilde{Q}(D)-Q(D)|\leq |Q(D)-Q(D')|+\mathrm{Err}_{\mathcal{M}}(D',\varepsilon_{\min}(D)/8,$ $\beta/2)$,
where the first term captures the error caused by truncation and the second term captures the error introduced by $\mathcal{M}$.
}
\end{proof}

As shown in the above theorem, the error consists of two components.
Component~I arises from excluding 
$\tilde{O}\!\left(1/\varepsilon_{\min}(D)\right)$ 
records during the doubly-truncation step. 
Under the DP setting, it has been shown~\cite{chen2025general} that any $\varepsilon$-DP mechanism at least incurs an error of $|Q(D) - Q(D')|$, where $D'$ is an arbitrary dataset obtained from $D$ by removing $\tilde{O}\!\left(1/\varepsilon_{\min}(D)\right)$ records. Therefore, Component~I is at most 
$\mathrm{Err}_{\mathcal{M}}\!\left(D, \Omega(\varepsilon_{\min}(D)), \beta\right)$
by a logarithmic factor.

For Component~II, which arises from the intrinsic DP error of the downstream record-DP mechanism, since DP error is typically inversely proportional to the privacy budget, this term is $O(\mathrm{Err}_{\mathcal{M}}\!\left(D', \Omega(\varepsilon_{\min}(D)), \beta\right))$.
Moreover, since $D$ and $D'$ differ by only $\tilde{O}\!\left(1/\varepsilon_{\min}(D)\right)$ number of records, in most cases, 
\begin{equation*}
    O\big(\mathrm{Err}_{\mathcal{M}}\!\bigl(D', \Omega(\varepsilon_{\min}(D)), \beta)\bigr)\approx O\bigl(\mathrm{Err}_{\mathcal{M}}\!\left(D, \Omega(\varepsilon_{\min}(D)), \beta\right)\bigr).
\end{equation*}
Overall, the above error is bounded by $\tilde{O}(\mathrm{Err}_{\mathcal{M}}\!\left(D, \Omega(\varepsilon_{\min}(D)), \beta\right))$.
Below, we will show the error guarantees for several common tasks under our mechanism.

\paragraph{Sum estimation}
For sum estimation, we first denote $F(r)$ as the value of record $r$, which is defined over the range $[0,1,\ldots,B]$.
Under this notation, the most straightforward solution in record-DP is the Laplace mechanism, which achieves an error bound of $O(B/\varepsilon)$. 
Furthermore, the state-of-the-art (SOTA) summation estimator from~\cite{dong2023universal} achieves an error bound of $\tilde{O}\bigl(\max_{r \in D} F(r) / \varepsilon\bigr)$.
This instance-specific error is more desirable since $\max_{r \in D} F(r) \ll B$ for most datasets.
In our framework, Component~I in Theorem~\ref{eq:utility-bound} incurs an error of $\tilde{O}(\max_{r \in D} F(r)$ /$\varepsilon_{\min}(D))$.
\xinghe{For Component~II, invoking~\cite{dong2023universal} as the downstream record-DP mechanism with a privacy budget $\tau_{i^\star}/4\geq\varepsilon_{\min}(D)/8$ over $D'$ yields an error of 
$\tilde{O}(\max_{r \in D'} F(r)$ /$\varepsilon_{\min}(D))$.
Since $D' \subseteq D$, we have $\max_{r \in D'}$ $F(r) \leq $ $\max_{r \in D}$ $F(r)$. Therefore, it can be further bounded by $\tilde{O}(\max_{r \in D} F(r)$ /$\varepsilon_{\min}(D))$.
In conclusion, the overall error is 
$\tilde{O}($ $\max_{r \in D}$ $F(r)$ /$\varepsilon_{\min}(D))$,
which aligns with our target.}

\paragraph{Distinct count}
For distinct count, Component~I is $\tilde{O}\!\left(1/\varepsilon_{\min}(D)\right)$. 
\xinghe{
Meanwhile, using the Laplace mechanism (Lemma~\ref{lem:laplace}) as the record-DP downstream mechanism on $D'$ with a privacy budget $\tau_{i^\star}/4 \geq \varepsilon_{\min}(D)/8$ and a sensitivity of $1$, Component~II incurs an error of $O\!\left(1/\varepsilon_{\min}(D)\right)$.
Combining the two components yields a total error of  $\tilde{O}\!\left(1/\varepsilon_{\min}(D)\right)$, which is at most a doubly logarithmic factor larger than the corresponding record-DP error.}

\paragraph{Maximum estimation}
For maximum estimation, we use \textit{rank error} as the metric, which measures the true rank of the estimated maximum.
In record-DP, we can invoke~\cite{dong2023universal} to obtain a maximum with rank error $\tilde{O}\!\left(1/\varepsilon_{\min}(D)\right)$. 
In our mechanism, using~\cite{dong2023universal} as the base DP mechanism, Component~I yields a rank error of $\tilde{O}\!\left(1/\varepsilon_{\min}(D)\right)$, while Component~II also contributes an error of $\tilde{O}\!\left(1/\varepsilon_{\min}(D)\right)$. 
Therefore, the overall rank error remains $\tilde{O}\!\left(1/\varepsilon_{\min}(D)\right)$, as desired.

\paragraph{Queries over relational databases}
\xinghe{A relational database with foreign-key constraints can be formulated as a user--record dataset, where tuples in the referenced relations represent users and tuples in the referencing relations represent records.
The foreign keys thus encode the ownership relationships between users and records.
However, in schemas with associative relations, these constraints may induce complex ownership relationships in which each record is jointly associated with multiple users.
Such ownership structures are more complex than those supported by our framework, which therefore does not currently apply to relational databases with such constraints.
}

%% file: main_text/7_experiment.tex
\section{Experiments}\label{sec:exp}

\bbd{In this section, we present empirical results for our proposed Hybrid-PDP framework under a variety of privacy specifications and across both real-world and synthetic settings.
We evaluate the following three fundamental queries. 
For \textit{Count}, we incorporate the classic Laplace mechanism (Lemma~\ref{lem:laplace}) into our framework. 
For \textit{Sum}, we integrate the SOTA Laplace-based, instance-specific standard DP \textit{Sum} mechanism~\cite{dong2023universal}. 
For \textit{Max}, we incorporate the Laplace-based exponential mechanism from~\cite{dong2023universal}, which requires the applicability condition $|D| > \tilde{O}(1/\varepsilon)$ and outputs $-1$ otherwise.}
For evaluation metrics, we measure errors using relative error (RE) for \textit{Count} and \textit{Sum}, and use the more meaningful relative rank error for \textit{Max}.

\begin{table*}[h]
    \caption{\gbd{Comparison of different mechanisms under Hybrid-PDP for Count, Sum, and Max. RE (\%) denotes relative error.}}
    \label{tab:main_results}

    \definecolor{mygray}{HTML}{E7E7E7}

    \setlength{\arrayrulewidth}{0.45pt}
    \setlength{\doublerulesep}{0.55pt}
    \renewcommand{\arraystretch}{0.95}
    \setlength{\tabcolsep}{5.0pt}

    \newcommand{\VC}[1]{%
        \begin{tabular}[c]{@{}c@{}}
            #1
        \end{tabular}%
    }
    \newcommand{\TopDbl}{\hhline{===||===||==||==||==}}
    \newcommand{\MidDbl}{\hhline{===||===||==||==||==}}

    \centering
    \resizebox{1.72\columnwidth}{!}{%
        \begin{tabular}{c|c|c||c|c|c||c|c||c|c||c|c}
            \TopDbl

            \multicolumn{3}{c||}{Dataset}
            & CK Pay
            & NY Salary
            & VLE
            & Online Retail
            & Czech Bank
            & TPCH Order
            & TPCH Lineitem
            & IBM
            & MultiSyn \\
            \MidDbl

            \multicolumn{3}{c||}{\VC{Privacy Specification}}
            & \multicolumn{3}{c||}{\VC{$\Phi_{\mathrm{rec}}(r)\propto 2^{\mathcal{N}(0,\sigma^2)}$}}
            & \multicolumn{2}{c||}{\VC{$\Phi_{\mathrm{rec}}(r)\propto F(r)\cdot 2^{\mathcal{N}(0,\sigma^2)}$}}
            & \multicolumn{2}{c||}{\VC{$\Phi_{\mathrm{rec}}(r)\propto \sqrt{F(r)} \cdot 2^{\mathcal{N}(0,\sigma^2)}$}}
            & \multicolumn{2}{c}{\VC{Multi-source}} \\
            \MidDbl


            \multirow{10}{*}{Count}
            & \multirow{2}{*}{Naive}
            & \multirow{2}{*}{\VC{RE (\%)\\Time (ms)}}
            & $>$100
            & $>$100
            & $>$100
            & $>$100
            & $>$100
            & $>$100
            & $>$100
            & $>$100
            & $>$100 \\
            & &
            & 0.32
            & 1.99
            & 55.9
            & 1.03
            & 3.66
            & 3.52
            & 27.3
            & 2.53
            & 10.7 \\
            \hhline{~--||---||--||--||--}

            & \multirow{2}{*}{Folded}
            & \multirow{2}{*}{\VC{RE (\%)\\Time (s)}}
            & 91.0
            & 9.08
            & 2.13
            & 49.9
            & \cellcolor{mygray}5.56
            & 2.71
            & 2.33
            & 15.52
            & 2.26 \\
            & &
            & 0.37
            & 1.93
            & 26.8
            & 0.59
            & 3.26
            & 5.68
            & 23.7
            & 4.85
            & 5.51 \\
            \hhline{~--||---||--||--||--}

            & \multirow{2}{*}{Oracle-mean}
            & \multirow{2}{*}{\VC{RE (\%)\\Time (s)}}
            & 3.64
            & 3.56
            & 70.8
            & 28.1
            & 51.7
            & 3.79
            & 67.4
            & 22.2
            & 6.02 \\
            & &
            & 0.55
            & 4.15
            & 30.7
            & 0.72
            & 5.13
            & 13.5
            & 29.4
            & 5.85
            & 9.01 \\
            \hhline{~--||---||--||--||--}

            & \multirow{2}{*}{Oracle-median}
            & \multirow{2}{*}{\VC{RE (\%)\\Time (s)}}
            & 0.348
            & 0.1113
            & 72.5
            & 18.54
            & 55.0
            & 1.127
            & 65.8
            & 11.51
            & 1.540 \\
            & &
            & 0.61
            & 4.30
            & 33.2
            & 0.87
            & 5.21
            & 15.3
            & 35.2
            & 12.3
            & 9.00 \\
            \hhline{~--||---||--||--||--}

            & \multirow{2}{*}{Framework}
            & \multirow{2}{*}{\VC{RE (\%)\\Time (s)}}
            & \cellcolor{mygray}0.00302
            & \cellcolor{mygray}0.01116
            & \cellcolor{mygray}1.922
            & \cellcolor{mygray}5.40
            & 16.35
            & \cellcolor{mygray}0.817
            & \cellcolor{mygray}0.881
            & \cellcolor{mygray}2.97
            & \cellcolor{mygray}0.527 \\
            & &
            & 0.61
            & 2.67
            & 24.9
            & 1.02
            & 3.01
            & 5.31
            & 20.1
            & 2.53
            & 7.83 \\
            \MidDbl


            \multirow{12}{*}{Sum}
            & \multirow{2}{*}{Naive}
            & \multirow{2}{*}{\VC{RE (\%)\\Time (ms)}}
            & 100
            & 100
            & 100
            & 100
            & 100
            & 100
            & 100
            & 100
            & 100 \\
            & &
            & 17.1
            & 62.8
            & 819
            & 40.3
            & 78.8
            & 108
            & 457
            & 67.9
            & 224 \\
            \hhline{~--||---||--||--||--}

            & \multirow{2}{*}{Folded}
            & \multirow{2}{*}{\VC{RE (\%)\\Time (s)}}
            & 100
            & 100
            & 20.9
            & 100
            & 100
            & 100
            & 17.96
            & 87.3
            & 10.01 \\
            & &
            & 0.37
            & 2.32
            & 30.1
            & 0.63
            & 4.30
            & 8.98
            & 43.7
            & 2.86
            & 6.70 \\
            \hhline{~--||---||--||--||--}

            & \multirow{2}{*}{Folded-sum}
            & \multirow{2}{*}{\VC{RE (\%)\\Time (s)}}
            & 100
            & 100
            & 50.8
            & 100
            & 100
            & 100
            & 94.3
            & $>$100
            & 99.4 \\
            & &
            & 0.68
            & 2.83
            & 33.1
            & 1.41
            & 3.66
            & 6.27
            & 21.6
            & 3.80
            & 8.77 \\
            \hhline{~--||---||--||--||--}

            & \multirow{2}{*}{Oracle-mean}
            & \multirow{2}{*}{\VC{RE (\%)\\Time (s)}}
            & 19.33
            & 7.11
            & 65.5
            & 35.9
            & 45.2
            & 2.68
            & 62.3
            & 30.8
            & 5.00 \\
            & &
            & 1.30
            & 7.66
            & 53.0
            & 1.80
            & 11.3
            & 44.4
            & 121
            & 10.5
            & 22.7 \\
            \hhline{~--||---||--||--||--}

            & \multirow{2}{*}{Oracle-median}
            & \multirow{2}{*}{\VC{RE (\%)\\Time (s)}}
            & 10.41
            & 1.785
            & 69.0
            & 33.9
            & 42.4
            & 0.899
            & 70.2
            & 29.5
            & 1.498 \\
            & &
            & 1.42
            & 8.57
            & 56.4
            & 2.07
            & 12.9
            & 46.1
            & 123
            & 30.0
            & 22.5 \\
            \hhline{~--||---||--||--||--}

            & \multirow{2}{*}{Framework}
            & \multirow{2}{*}{\VC{RE (\%)\\Time (s)}}
            & \cellcolor{mygray}0.1813
            & \cellcolor{mygray}0.1260
            & \cellcolor{mygray}3.16
            & \cellcolor{mygray}14.31
            & \cellcolor{mygray}2.11
            & \cellcolor{mygray}0.0378
            & \cellcolor{mygray}0.970
            & \cellcolor{mygray}2.17
            & \cellcolor{mygray}0.1062 \\
            & &
            & 0.71
            & 3.14
            & 31.5
            & 1.20
            & 4.19
            & 8.69
            & 42.1
            & 7.00
            & 9.30 \\
            \MidDbl


            \multirow{10}{*}{Max}
            & \multirow{2}{*}{Naive}
            & \multirow{2}{*}{\VC{RE (\%)\\Time (ms)}}
            & 100
            & 100
            & 100
            & 100
            & 100
            & 100
            & 100
            & 100
            & 100 \\
            & &
            & 9.03
            & 37.6
            & 477
            & 21.6
            & 43.4
            & 61.6
            & 219
            & 35.8
            & 127 \\
            \hhline{~--||---||--||--||--}

            & \multirow{2}{*}{Folded}
            & \multirow{2}{*}{\VC{RE (\%)\\Time (s)}}
            & 100
            & 100
            & 44.8
            & 100
            & 100
            & 100
            & 19.19
            & 100
            & 77.1 \\
            & &
            & 0.36
            & 2.03
            & 27.5
            & 0.59
            & 3.47
            & 6.34
            & 27.1
            & 2.66
            & 5.96 \\
            \hhline{~--||---||--||--||--}

            & \multirow{2}{*}{Oracle-mean}
            & \multirow{2}{*}{\VC{RE (\%)\\Time (s)}}
            & 41.9
            & 4.00
            & 1.484
            & 2.94
            & 4.12
            & 1.766
            & 1.504
            & 8.20
            & 0.846 \\
            & &
            & 0.90
            & 5.09
            & 37.6
            & 1.16
            & 6.10
            & 18.7
            & 43.4
            & 6.86
            & 15.0 \\
            \hhline{~--||---||--||--||--}

            & \multirow{2}{*}{Oracle-median}
            & \multirow{2}{*}{\VC{RE (\%)\\Time (s)}}
            & 34.7
            & 5.06
            & 1.364
            & 4.43
            & 3.68
            & 1.475
            & 1.473
            & 4.23
            & 0.808 \\
            & &
            & 0.99
            & 5.54
            & 38.0
            & 1.33
            & 6.35
            & 20.1
            & 44.0
            & 7.02
            & 15.2 \\
            \hhline{~--||---||--||--||--}

            & \multirow{2}{*}{Framework}
            & \multirow{2}{*}{\VC{RE (\%)\\Time (s)}}
            & \cellcolor{mygray}0.469
            & \cellcolor{mygray}0.1258
            & \cellcolor{mygray}1.352
            & \cellcolor{mygray}1.644
            & \cellcolor{mygray}1.707
            & \cellcolor{mygray}0.1447
            & \cellcolor{mygray}0.594
            & \cellcolor{mygray}0.626
            & \cellcolor{mygray}0.1469 \\
            & &
            & 0.66
            & 2.88
            & 24.0
            & 1.05
            & 3.19
            & 5.93
            & 22.9
            & 2.68
            & 8.48 \\

            \TopDbl
        \end{tabular}%
    }
\end{table*}

\subsection{Setup}\label{sec:setup}
\paragraph{Datasets.}

\bbd{We consider datasets from different real-world settings, including banking transactions (Czech Bank~\cite{berka_dataset}), salary payments (CK Pay~\cite{employee_payroll_cook_county}, NY Salary~\cite{salary_information_for_state_authorities}), retail statistics (Online Retail~\cite{online_retail_ii_uci}), web user behavior data (VLE~\cite{open_university_learning_analytics_uci}), and the standard benchmark databases (TPCH Order and TPCH Lineitem~\cite{tpch_homepage}).
We further consider two multi-source datasets, i.e., IBM, the \gbd{simulator-generated} version 2.1 example dataset released as part of the AMLSim benchmark~\cite{amlsim_dataset}, which contains transactions from multiple banks, and MultiSyn, which is constructed by aggregating three synthetic datasets to simulate data collected from multiple sources.
In total, we use five real-world datasets, three benchmark datasets, and a suite of synthetic datasets, covering diverse applications, sizes and distributions, as shown in Table~\ref{tab:statistics}.
Common preprocessing practices, such as excluding abnormal values, are applied to all datasets. 
For more information on dataset-specific preprocessing and further details of the synthetic datasets, please refer to Appendix~\ref{preprocess}.}

\paragraph{Privacy Budgets.}
We adopt log-scale randomization for both user- and record-level privacy budgets, as it yields greater heterogeneity and better captures diverse privacy requirements. This setting is well accepted in the DP research community~\cite{dong2022r2t,zhang2012functional}.
For user-level privacy budgets, all real-world datasets share the same privacy specification. 
\bbd{Specifically, for each user or user group, defined by certain attributes in the original databases, the privacy budget is independently sampled according to $\Phi_{\mathrm{user}}(u) \propto 2^{\mathcal{N}(0,\sigma^2)}$.
For record-level privacy budgets, we consider three different privacy budget functions to demonstrate the generalizability of our framework.}
\bbd{For the CK Pay, NY Salary, and VLE datasets, we use the log-normal function
$\Phi_{\mathrm{rec}}(r) \propto 2^{\mathcal{N}(0,\sigma^2)}$,
independent of record values.
For the remaining datasets, we introduce positive but nondeterministic correlations between record values and budgets.
Specifically, we use the linear function
$\Phi_{\mathrm{rec}}(r) \propto 2^{\mathcal{N}(0,\sigma^2)} \cdot F(r)$
for Online Retail and Czech Bank, and the square-root function
$\Phi_{\mathrm{rec}}(r) \propto 2^{\mathcal{N}(0,\sigma^2)} \cdot \bigl(F(r)\bigr)^{0.5}$
for TPCH Order and TPCH Lineitem.
For IBM and MultiSyn, which aggregate multiple data sources, we assign different record-level privacy functions across sources using the three types above, consistent with the multi-bank scenario in Section~\ref{sec:intro}.
Further details are provided in Appendix~\ref{privacyspe}.}

\paragraph{Other Parameters}
All the empirical results are obtained using a Linux server equipped with dual AMD EPYC 9555 64-Core Processors and 2~TB of memory.
For the global bounds on user- and record-level budgets, we set $\check{\varepsilon}_\mathrm{user} = 2^{-10}$, $\check{\varepsilon}_\mathrm{rec} = 2^{-20}$, $\hat{\varepsilon}_\mathrm{user} = \hat{\varepsilon}_\mathrm{rec} = 2^{4}$, and $\hat{m} = 2^{16}$.
We set the failure rate to $\beta = 10\%$ across all experiments. For each setting, we repeat the experiment 50 times, drop the top and bottom 10 results, and report the average over the remaining $30$ runs for both RE and running time. This 40\% trimming is more conservative than the failure rate and helps ensure reliability.

\paragraph{\bbd{Baselines.}}

\bbd{
To highlight the advantages of our framework and assess the importance of its components, we compare it against \gbd{5} baselines. 
\gbd{Specifically, \textbf{Naive} demonstrates the limitation of relying on conservative global bounds, while \textbf{Oracle-mean} and \textbf{Oracle-median} show that selecting an appropriate threshold is essential for high utility. }
Additionally, for direct comparison with SOTA PDP solutions~\cite{chen2025general,dajun2024PDP}, \textbf{Folded} and \textbf{Folded-sum} use a global bound $\hat{m}$ to reduce Hybrid-PDP to record-PDP.
Due to space limitations, details of these baselines are deferred to Appendix~\ref{app:baseline}.}

\subsection{Results}
\bbd{Due to space limitations, we present only the main experimental results 
in this section and defer the details to Appendix~\ref{app:exp}.}

\subsubsection{Utility and Efficiency.} \label{sec:mainexp}

\bbd{
Table~\ref{tab:main_results} evaluates the utility and runtime of different mechanisms for \textit{Count}, \textit{Sum}, and \textit{Max} across nine datasets under diverse privacy specifications. 
Overall, our framework consistently achieves high utility across different queries, datasets, and privacy settings: it obtains the lowest RE in most \gbd{(26 out of 27)} dataset-query combinations, with the only exception being \textit{Count} on Czech Bank, and achieves an average reduction of \gbd{$16.56\times$} and a median reduction of \gbd{$6.61\times$} in RE compared with the best baseline in each combination, while requiring at most \gbd{$2.45\times$} the runtime of any baseline except \textbf{Naive}.
Notably, in the multi-source setting corresponding to the multi-party scenario in Section~\ref{sec:intro}, our framework consistently maintains an RE below $3\%$ and significantly outperforms all baselines. 
These results demonstrate the robustness and efficiency of our framework in achieving Hybrid-PDP under heterogeneous privacy requirements.
}

\subsubsection{Privacy Budget Skewness}\label{sec:skewness}
\bbd{
To evaluate robustness to privacy-budget skewness, we consider \textit{Sum} on synthetic datasets under three record-level privacy specifications and vary the standard deviation $\sigma$ from $2^{-5}$ to $2^{2}$, where larger $\sigma$ induces greater budget skewness. 
As shown in Figure~\ref{fig:skewness} in Appendix~\ref{app:skewness},
our framework consistently achieves the best utility across all settings and maintains an RE below \textbf{$20\%$} even under highly skewed privacy-budget distributions. 
In contrast, the performance of the baselines degrades substantially with increasing skewness or consistently incurs large errors.
Notably, for $\sigma \in \{2^{-5},\ldots,2^{-1}\}$, our framework achieves at least $10\times$ lower RE than all the baselines. 
These results demonstrate the consistently strong performance of our framework under varying privacy-budget skewness.
}

\subsubsection{User-record Ownership Distribution}
\bbd{
To study the impact of user-record ownership, we evaluate \textit{Sum} under Poisson ownership distributions, i.e., $\Pr(m(D,u)=k)=\lambda^k e^{-\lambda}/k!$, varying $\lambda$ from $2^{0}$ to $2^{7}$, where larger $\lambda$ corresponds to more records per user on average. 
We consider the same three privacy specifications as in Section~\ref{sec:skewness}, with $\sigma=2^{-1}$. 
As shown in Figure~\ref{fig:lambda} in Appendix~\ref{app:ownership}, 
our framework consistently achieves the best utility across different ownership distributions, with REs below $3\%$ in all settings. 
In contrast, the baselines either deteriorate as the expected number of records per user increases or consistently incur large errors. 
Notably, at $\lambda=2^7$, our framework maintains REs below \textbf{$1\%$}, while all the baselines exceed \textbf{$24\%$} across all three privacy specifications. 
These results demonstrate the effectiveness of our framework under varying user-record ownership distributions.
}

\subsubsection{Optimality of $\tau$.}\label{sec:optimal}

\bbd{
To evaluate the effectiveness of our threshold selection, we manually vary $\tau$ from $2^{-9}$ to $2^{-2}$ and compare the resulting REs with those obtained by our framework on four datasets for \textit{Sum}. 
As shown in Table~\ref{tab:tau} in Appendix~\ref{app:optimal}, 
performance is highly sensitive to the choice of $\tau$ and varies significantly across datasets: an inappropriate value just $2\times$ larger than the optimal can lead to more than a \textbf{$500\times$} increase in RE. 
This not only underscores the importance of selecting an appropriate $\tau$ for data utility, but also demonstrates that a single predefined global threshold cannot consistently balance truncation error and DP noise across different datasets. 
Importantly, despite allocating a portion of the privacy budget to privately estimating $\tau$, our framework consistently achieves REs within a small constant factor of those obtained with the best manually tuned, non-private $\tau$. 
Overall, these results show that our framework can effectively select an appropriate $\tau$ for each case to achieve high data utility.
}

\subsubsection{Privacy Allocation}\label{sec:allocation}

\bbd{
We investigate how privacy-budget allocation between the two stages of Algorithm~\ref{alg:Hybrid-PDP} affects utility. We introduce a parameter $\alpha \in(0\%,100\%)$, where an $\alpha$ fraction of the privacy budget is allocated to threshold selection and the remaining $(100\%-\alpha)$ fraction is allocated to the downstream query mechanism
(see Appendix~\ref{app:allocation} for the details). 
We evaluate $\alpha \in \{10\%, 30\%$, $ 50\%$, $ 70\%$, $90\%\}$ on four datasets across all three queries. 
As shown in Table~\ref{tab:alpha_results}, 
the preferred allocation varies across queries and datasets: \textit{Count} generally favors larger values of $\alpha$, while \textit{Sum} and \textit{Max} tend to favor smaller values, and no single allocation consistently achieves the best utility. Importantly, the equal allocation $\alpha=50\%$ adopted in our framework performs consistently well, with RE within \textbf{$3\times$} of the best-performing allocation in all 12 settings. Since selecting a dataset-specific optimal allocation would 
require additional tuning and 
incur extra privacy cost, these results support using $\alpha=50\%$ as the default in our framework.
}

%% file: main_text/8_conclusion.tex
\section{CONCLUSION}\label{sec:conclusion}
In this work, we propose Hybrid-PDP, a novel DP model that provides personalized privacy guarantees simultaneously at both the record- and user-levels.
\bbd{We develop a general framework that reduces the Hybrid-PDP problem to a record-DP problem, enabling the use of any record-DP mechanism to achieve Hybrid-PDP protection.
Several directions remain for future work.
 One is to extend Hybrid-PDP to dynamic settings, where streams of users and records arrive over time. 
Another is to integrate Hybrid-PDP with approximate $(\epsilon,\delta)$-DP, where each user/record is associated with both a privacy budget and a failure probability. In such a setting, the record-level stability and dominance lemmas remain valid, while threshold-selection needs to jointly account for $\epsilon$ and $\delta$. 
Moreover, advanced composition~\cite{dwork2014algorithmic} can tighten the privacy bounds across multiple selection iterations.
Another direction is to study scenarios in which a record is jointly owned by multiple users~\cite{dajun2024PDP}.
In relational databases, this corresponds to associative relations.}

%% file: appendix.tex
\appendix

\bbd{\section{Artifact Availability}
We make all artifacts needed to evaluate and reproduce our results, including implementation details, scripts, and environment configurations, available at~\url{https://github.com/Saddlerdroph/Hybrid-PDP}.}

\section{Dataset Statistics and Preprocessing Techniques}
\label{app:data}

\begin{table*}[htbp]
  \caption{Dataset statistics. Here, med denotes the median, and max denotes the maximum value.}
  \label{tab:statistics}
  \setlength{\arrayrulewidth}{0.45pt}
  \setlength{\doublerulesep}{0.55pt}
  \renewcommand{\arraystretch}{1.2}
  \setlength{\tabcolsep}{5.0pt}

  \centering
  \footnotesize
  \makebox[\textwidth][c]{%
  \resizebox{0.85\textwidth}{!}{%
  \begin{tabular}{c|c|c|c|c|c|c}
    \hhline{=======}
    Dataset &
    \# of users &
    \# of records &
    $\mathrm{med}_{u\in D}(m(D,u))$ &
    $\mathrm{max}_{u\in D}(m(D,u))$ &
    $\mathrm{med}_{r\in D}(F(r))$ &
    $\mathrm{max}_{r\in D}(F(r))$ \\
    \hhline{=======}

    CK Pay &
    $\sim5.3 \times 10^{4}$ &
    $\sim2.3 \times 10^{5}$ &
    2 &
    14 &
    $\sim1.6 \times 10^{4}$ &
    $\sim2.4 \times 10^{5}$ \\
    \hhline{-------}

    NY Salary &
    $\sim2.1 \times 10^{5}$ &
    $\sim8.9 \times 10^{5}$ &
    4 &
    59 &
    $\sim7.5 \times 10^{4}$ &
    $\sim2.0 \times 10^{6}$ \\
    \hhline{-------}

    VLE &
    $\sim2.6 \times 10^{4}$ &
    $\sim1.1 \times 10^{7}$ &
    270 &
    6389 &
    2 &
    $\sim7.0 \times 10^{3}$ \\
    \hhline{-------}

    Online Retail &
    $\sim2.6 \times 10^{4}$ &
    $\sim5.4 \times 10^{5}$ &
     10 &
    1114 &
    $10$ &
    $\sim1.7 \times 10^{5}$ \\
    \hhline{-------}

    Czech Bank &
    $\sim4.5 \times 10^{3}$ &
    $\sim1.1 \times 10^{6}$ &
    12 &
    675 &
    12 &
    $\sim8.7 \times 10^{4}$ \\
    \hhline{-------}

    TPCH Order &
    $\sim1.0 \times 10^{5}$ &
    $\sim1.5 \times 10^{6}$ &
    14 &
    41 &
    $\sim1.4 \times 10^{5}$ &
    $\sim5.6 \times 10^{5}$ \\
    \hhline{-------}

    TPCH Lineitem &
    $\sim1.0 \times 10^{4}$ &
    $\sim6.0 \times 10^{6}$ &
    600 &
    694 &
    $\sim9.5 \times 10^{5}$ &
    $\sim5.2 \times 10^{6}$ \\
    \hhline{-------}

    \bbd{IBM} &
    \bbd{$\sim2.0 \times 10^{4}$} &
    \bbd{$\sim8.9 \times 10^{5}$} &
    \bbd{45} &
    \bbd{86} &
    \bbd{51} &
    \bbd{$\sim3.0 \times 10^{3}$} \\
    \hhline{-------}
    
    \bbd{MultiSyn} &
    \bbd{$\sim1.9 \times 10^{5}$} &
    \bbd{$\sim3.0 \times 10^{6}$} &
    \bbd{16} &
    \bbd{36} &
    \bbd{$\sim1.3 \times 10^{6}$} &
    \bbd{$\sim6.3 \times 10^{6}$} \\
    \hhline{-------}

    \end{tabular}%
  }%
  }

\end{table*}

\subsection{Dataset Statistics, Preprocessing and Definitions}\label{preprocess}

\bbd{\paragraph{Statistics}
We summarize the key statistics of all datasets used in our experiments, including the numbers of users and records as well as their ownership and value distributions, in Table~\ref{tab:statistics}.}

\paragraph{Preprocessing.}

\begin{table}[htbp]
  \definecolor{mygray}{HTML}{E7E7E7}
  \setlength{\arrayrulewidth}{0.45pt}
  \setlength{\doublerulesep}{0.55pt}
  \renewcommand{\arraystretch}{1.3}
  \setlength{\tabcolsep}{6.0pt}

  \newcommand{\VC}[1]{\begin{tabular}[c]{@{}c@{}}#1\end{tabular}}
  \newcommand{\TopDblIII}{\hhline{===}}
  \newcommand{\MidDblIII}{\hhline{===}}
  \caption{Definitions of users, records, and values across datasets.}
  \centering
  \footnotesize
  \resizebox{1\columnwidth}{!}{%
    \begin{tabular}{c|c|c}
      \TopDblIII
      Dataset Name  & User Identifying Attributes                 & Value Attributes                  \\
      \MidDblIII
      CK Pay        & Bureau, employee      & Quarterly salary (USD)      \\
      \hhline{---}
      NY Salary     & Authority, employee   & Annual salary (USD)         \\
      \hhline{---}
      VLE           & Student               & Sum click (Click)            \\
      \hhline{---}
      Online Retail & Invoice               & Price $\times$ quantity (GBP)\\
      \hhline{---}
      Czech Bank    & Account               & Amount (CZK)                 \\
      \hhline{---}
      TPCH Order    & Customer              & Total price                  \\
      \hhline{---}
      TPCH Lineitem & Supplier              & Extended price                 \\
      \hhline{---}
      IBM
      & \gbd{Originating account}
      & Base amount (USD) \\
        \hhline{---}
      MultiSyn
      & Synthetic User ID
      & Synthetic Value \\
 \TopDblIII
    \end{tabular}%
  }

\label{Definition_dataset}
\end{table}

For all raw datasets, we first remove 
(i) rows with missing target values and 
(ii) rows whose user-identifying attributes are missing.
\bbd{We quantize fractional record values to integers and clip all record values to the range $[0,2^{30}]$.}
Using the integer domain is not a limitation but a standard choice in the DP literature, since fractional values can be quantized to integers.

\paragraph{Definitions of users, records, and values.}
In our experiments, each retained row after preprocessing corresponds to a \textit{record}. 
The \textit{value} of a record is the numeric attribute used by the query.
Meanwhile, a \textit{user} is defined as the collection of records that share the same set of user-identifying attributes in the retained dataset.
The concrete definitions of record value attributes and user-identifying attributes for each dataset except MultiSyn are summarized in Table~\ref{Definition_dataset}.
\gbd{Notably, in IBM dataset, each transaction involves an originating account and a beneficiary account. Since our analysis focuses on the transactions initiated by each account, we assign each transaction record to its originating account, while retaining the beneficiary account as a transaction attribute.}
\bbd{
For MultiSyn, we independently generate $10^6$ records from each of three sources, with values sampled from a Gaussian distribution with mean and standard deviation $2^{20}$. Values outside $[0,2^{30}]$ are discarded and resampled until they fall within the range, after which they are rounded to integers.
We then independently sample each user's record count from
$\Pr(m(D,u)=k)=\lambda^k e^{-\lambda}/k!$ with $\lambda=2^4$, until the cumulative count reaches $3\times10^6$.
Finally, we randomly shuffle the $3\times10^6$ records from the three sources
and partition them among users according to the sampled counts.}

\subsection{Privacy Specification}\label{privacyspe}

\begin{table}[hbtp]
\definecolor{mygray}{HTML}{E7E7E7}
\setlength{\arrayrulewidth}{0.45pt}
\setlength{\doublerulesep}{0.55pt}
\renewcommand{\arraystretch}{1.3}
\setlength{\tabcolsep}{6.0pt}
\caption{Grouping schemes and dataset-specific parameters for personalized privacy budgets.}
\centering
\footnotesize
\resizebox{1\columnwidth}{!}{%
\begin{tabular}{c|c|c|c|c}
\hline\hline
Category & Dataset Name & User Group & Record Group & Parameter \\
\hline\hline

\multirow{3}{*}{Normal}
  & CK Pay      & Bureau    & Season & \multirow{3}{*}{$\mu=-2$} \\
\cline{2-4}
  & NY Salary   & Authority & Month  & \\
\cline{2-4}
  & VLE         & Student   & Site   & \\

\hline

\multirow{2}{*}{Linear}
  & Online Retail & Invoice & \multirow{4}{*}{\shortstack{No group}}
  & $\theta=-5$ \\
\cline{2-3}\cline{5-5}
  & Czech Bank    & Account & & $\theta=-12$ \\

\cline{1-3}\cline{5-5}

\multirow{2}{*}{Sqrt}
  & TPCH Order    & Area     & & $\gamma=-12$ \\
\cline{2-3}\cline{5-5}
  & TPCH Lineitem & Supplier & & $\gamma=-12$ \\

\hline\hline
\end{tabular}%
}

\label{privacy_speci}
\end{table}

\begin{table}[hbtp]

\definecolor{mygray}{HTML}{E7E7E7}
\setlength{\arrayrulewidth}{0.45pt}
\setlength{\doublerulesep}{0.55pt}
\renewcommand{\arraystretch}{1.3}
\setlength{\tabcolsep}{6.0pt}

\arrayrulecolor{black} 

\caption{Source-specific parameters for multi-source datasets.}
\centering
\footnotesize
\resizebox{1\columnwidth}{!}{%
\begin{tabular}{c|c|c|c}
\hline\hline
Dataset Name & Source & Category & Parameter \\
\hline\hline

\multirow{4}{*}{IBM}
  & Intra-bank (Bank A)     & Linear & $\theta=-7$ \\
\cline{2-4}
  & Intra-bank (Bank B)     & Linear & $\theta=-8$ \\
\cline{2-4}
  & Intra-bank (Bank C)     & Linear & $\theta=-9$ \\
\cline{2-4}
  & Inter-bank (Shared)     & Normal & $\mu=-2$ \\

\hline

\multirow{3}{*}{MultiSyn}
  & Source 1 & Normal & $\mu=-2$ \\
\cline{2-4}
  & Source 2 & Linear & $\theta=-20$ \\
\cline{2-4}
  & Source 3 & Sqrt   & $\gamma=-10$ \\

\hline\hline
\end{tabular}%
}

\label{tab:multi_source_params}
\end{table}

\paragraph{User-level budgets.} 
\bbd{For all datasets except the two multi-source datasets, we partition users into disjoint groups according to the \textit{user group} attribute listed in Table~\ref{privacy_speci}.
For the two multi-source datasets, each user is treated as an individual group.}
Users within the same group share the same privacy budget, while the budgets of different groups are independently sampled from a distribution, aligning with common real-world privacy requirements.
Specifically, for each user group, we sample a group-level budget $\varepsilon = 2^{2} \cdot 2^{\mathcal{N}(0,\sigma^2)}$ with $\sigma=2^{-1}$, and assign it to every user $u$ in the group. 
To enforce the global bounds, if the sampled budget falls outside $[\check{\varepsilon}_\mathrm{user},\hat{\varepsilon}_\mathrm{user}]$, it is discarded and resampled until it satisfies the constraints.

\paragraph{Record-level budgets.}
Record-level privacy budgets are generated under three specifications, namely \textit{normal}, \textit{linear}, and \textit{sqrt}, in order to capture the heterogeneity of record-level privacy requirements in practice, where the budgets can be independent of the record value or scale with the record value.
Recall that $F(r)$ denotes the value associated with record $r$ as defined in Section~\ref{preprocess}.
\bbd{Unless otherwise specified, we use $2^{\mathcal{N}(0,\sigma^2)}$ with $\sigma = 0.5$ to reflect various privacy needs. 
Details of each privacy specification are listed below: }

\begin{itemize}
  \item \textit{Normal.}
  We partition records into disjoint record groups, and all records within the same group share a common record-level budget.
  For each record group, we sample  $\Phi_{\mathrm{rec}}(r)=2^{\mu}\cdot 2^{\mathcal{N}(0,\sigma^2)}$,
  where $\mu$ is fixed to $-2$ for all datasets under the normal specification.

  \item \textit{Linear.}
  We assign each record $r$ a budget proportional to its value: $\Phi_{\mathrm{rec}}(r)=2^{\theta}\cdot 2^{\mathcal{N}(0,\sigma^2)}\cdot F(r)$, where $\theta$ is dataset-specific as reported in Table~\ref{privacy_speci}.

  \item \textit{Sqrt.}
  We assign each record $r$ a budget proportional to the square root of its value:
    $\Phi_{\mathrm{rec}}(r)=2^{\gamma}\cdot 2^{\mathcal{N}(0,\sigma^2)}\cdot \bigl(F(r)\bigr)^{1/2}$, where $\gamma$ is dataset-specific as reported in Table~\ref{privacy_speci}.
\end{itemize}
Similar to the user-level case, to enforce global constraints, we require $\Phi_{\mathrm{rec}}(r)\in[\check{\varepsilon}_{\mathrm{rec}},\hat{\varepsilon}_{\mathrm{rec}}]$ for every record $r$.
\bbd{Notably, for multi-source datasets, where each dataset consists of records from different data sources, we assign each source an independent record-level privacy function, with the corresponding source-specific settings and parameters reported in Table~\ref{tab:multi_source_params}.
}

\section{Baselines}\label{app:baseline}
\bbd{In this section, we provide detailed descriptions of the configurations of the baselines used in our evaluation.}

\bbd{(1) \textbf{Naive}
uses the conservative global privacy bound introduced in Section~\ref{sec:strawman} as a uniform record-level privacy budget and directly executes the corresponding record-DP mechanism.}

\bbd{(2) \textbf{Oracle-mean} first splits both the record- and user-level privacy budgets evenly across the $10$ runs, allocating $1/10$ of the original budgets to each run. It then uses an oracle to sample $10$ random thresholds uniformly on a logarithmic scale. For each run, the corresponding threshold is used to doubly-truncate the dataset and execute the downstream query. Finally, we discard the two largest and two smallest query outputs and report the mean of the remaining outputs.}

\bbd{(3) \textbf{Oracle-median} follows the same threshold generation, privacy allocation, and trimming procedures as \textbf{Oracle-mean}, but reports the median of the query outputs.}

\bbd{(4) \textbf{Folded} reduces the Hybrid-PDP requirements to record-PDP using the public upper bound $\hat{m}$\footnote{\gbd{We must use the public bound $\hat{m}$, as directly using the data-dependent maximum contribution $m_{\max}(D)$ would violate Hybrid-PDP guarantees. Moreover, under Hybrid-PDP constraints, obtaining an accurate estimate of $m_{\max}(D)$ is infeasible due to the bidirectional dependence between users and records, as discussed in Section~\ref{attempt}.}} on the number of records contributed by each user. Specifically, for each record $r\in D(u)$, we define its effective record-level privacy budget as $\Phi_{\mathrm{eff}}(r)=\min{\Phi_{\mathrm{rec}}(r),\Phi_{\mathrm{user}}(u)/\hat{m}}$. We then apply the record-PDP framework of ~\cite{chen2025general} to the resulting record-PDP problem.}

\bbd{(5) \textbf{Folded-sum} applies the same reduction based on $\Phi_{\mathrm{eff}}$ and then uses the tuple-level PDP summation mechanism in~\cite{dajun2024PDP}, which is equivalent to a record-PDP mechanism in our setting. Since this mechanism is specifically designed for summation, \textit{Folded-sum} is evaluated only for Sum queries.}

\section{Detailed Experimental Results and Analysis}\label{app:exp}

\subsection{Utility and Efficiency.} \label{app:mainexp}

\bbd{
Table~\ref{tab:main_results} compares the utility and runtime of different mechanisms for the three investigated queries across nine datasets.
We then analyze the performance of different baselines and the reasons behind their utility loss.}

\bbd{
The baselines directly relying on global bounds are generally limited by the overly conservative record-level budget induced by $\check{\varepsilon}_{\mathrm{rec}}$, $\check{\varepsilon}_{\mathrm{user}}$ and $\hat m$.
Specifically, \textbf{Naive} incurs errors of at least $100\%$ across all settings, as the worst-case global bound results in excessively small record-level privacy budgets.
Similarly, although \textbf{Folded}, which relies solely on the global bound $\hat m$, is competitive for \textit{Count} and even outperforms our framework on Czech Bank, where the record-level privacy budget is set relatively small, it performs poorly for \textit{Sum} and \textit{Max} on most datasets.
A similar trend is observed for \textbf{Folded-sum}.
This is because the downstream mechanisms for \textit{Sum} and \textit{Max} involve multiple budget-consuming components and thus require sufficiently large record-level privacy budgets.
In particular, for \textit{Max}, an overly small budget may additionally violate the applicability condition of the underlying exponential mechanism, which requires $|D| > \tilde{O}(1/\varepsilon)$.
}


\bbd{
Turning to the oracle baselines, \textbf{Oracle-mean} and \textbf{Oracle-median}, which empirically aggregate the outputs of multiple runs with randomized thresholds, our framework consistently outperforms both across all queries and datasets while generally requiring less runtime.
This suggests that our threshold-selection mechanism is more effective than spreading the privacy budget across multiple candidate thresholds and aggregating their outputs.
}

\bbd{
Overall, our framework consistently achieves high utility across different queries, datasets, and privacy settings: it obtains the lowest RE in most \gbd{(26 out of 27)} dataset-query combinations, with the only exception being \textit{Count} in Czech Bank, and achieves an average reduction of \gbd{$16.56\times$} and a median reduction of \gbd{$6.61\times$} in RE compared with the best baseline in each combination, while requiring at most \gbd{$2.45\times$} the runtime of any baseline except \textbf{Naive}.
These performance gains are particularly important for preserving high data utility under heterogeneous privacy requirements.
Collectively, these findings provide strong empirical evidence for the effectiveness of our doubly-truncation framework in efficiently achieving Hybrid-PDP while preserving high data utility.
}

\begin{figure*}[htb]
    \centering
    \includegraphics[width=\textwidth]{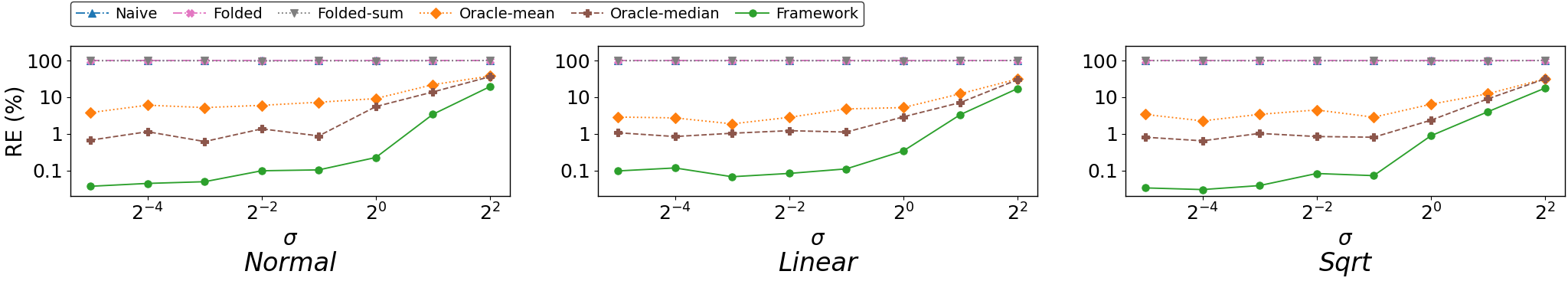}
    \caption{\gbd{Comparison of RE (\%) for \textit{Sum} under Hybrid-PDP across three privacy specifications and varying privacy budget skewness $\sigma$. Both axes are in log scale.}}
    \label{fig:skewness}
\end{figure*}

\subsection{Privacy Budget Skewness} 
\label{app:skewness}
To demonstrate the impact of privacy budget skewness, we report empirical results for \textit{Sum} on the synthetic datasets, under three privacy specifications. 
More specifically, we fix the user-level privacy budget function as $\Phi_\mathrm{user}(u)=2^{2}\cdot 2^{\mathcal{N}(0,\sigma^2)}$, and consider three record-level privacy budget functions: normal ($\Phi_\mathrm{rec}(r)=2^{-2}\cdot2^{\mathcal{N}(0,\sigma^2)}$), linear ($\Phi_\mathrm{rec}(r)=2^{-20}\cdot 2^{\mathcal{N}(0,\sigma^2)}\cdot F(r)$), and sqrt ($\Phi_\mathrm{rec}(r)=2^{-10}\cdot 2^{\mathcal{N}(0,\sigma^2)}\cdot (F(r))^{0.5}$). 
We vary the standard deviation $\sigma$ from $2^{-5}$ to $2^2$, with larger $\sigma$ corresponding to greater skewness.

\bbd{As shown in Figure~\ref{fig:skewness}, our framework consistently outperforms the baselines across all settings, maintaining an RE below $20\%$ even under the largest skewness considered, i.e., $\sigma=2^{2}$. 
In contrast, the three baselines that rely on global privacy-budget bounds, i.e., \textbf{Naive}, \textbf{Folded}, and \textbf{Folded-sum}, perform poorly across all settings, with REs close to $100\%$.
Among the two oracle baselines, \textbf{Oracle-mean} and \textbf{Oracle-median}, both retain reasonable utility when $\sigma$ is relatively small, whereas their REs increase to approximately $40\%$ as $\sigma$ grows, resulting in limited utility under highly skewed privacy budgets.
In comparison, our framework maintains a clear utility advantage over the oracle baselines across different levels of skewness.
In particular, for $\sigma \leq 2^{-1}$, our framework achieves at least $10\times$ lower RE than the oracle baselines, highlighting the effectiveness of our threshold-selection strategy in adapting to heterogeneous privacy budgets, compared with the random-threshold aggregation baselines.
These results demonstrate that our framework is substantially more robust to privacy-budget skewness than the considered baselines.
}
\begin{figure*}[hptb]
    \centering
    \includegraphics[width=\textwidth]{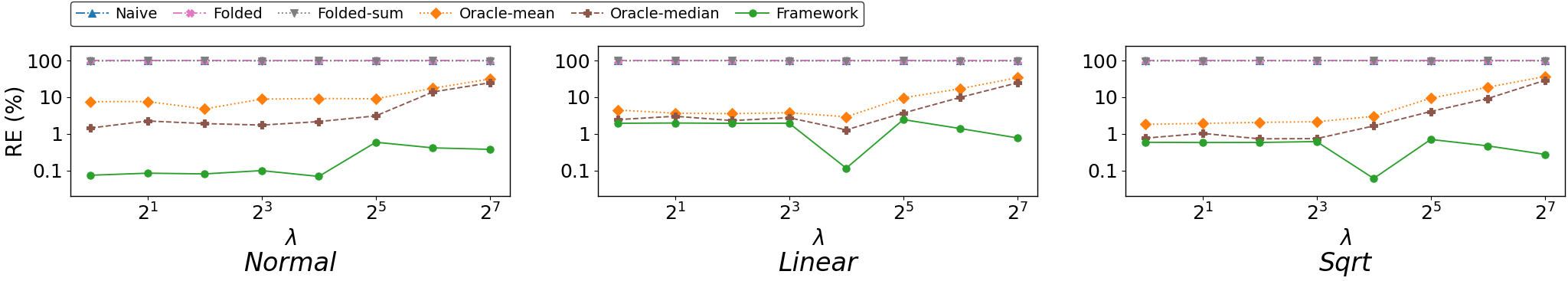}
    \caption{\gbd{Comparison of RE (\%) for \textit{Sum} under Hybrid-PDP across three privacy specifications and varying user-record ownership distributions. Both axes are in log scale.}}
    \label{fig:lambda}
\end{figure*}

\subsection{User-record Ownership Distribution}\label{app:ownership}
To study the impact of the user-record ownership distribution, we report empirical results for \textit{Sum} under different Poisson ownership distributions, $\Pr(m(D,u)=k) = \lambda^{k} e^{-\lambda}/k!$, varying $\lambda$ from $2^{0}$ to $2^{7}$, where larger $\lambda$ indicates a higher expected number of records owned by each user.
We also include the three privacy specifications considered in Section~\ref{sec:skewness}, with $\sigma=2^{-1}$ fixed to represent moderate privacy budget skewness.

\bbd{
Figure~\ref{fig:lambda} shows that our proposed framework achieves the best data utility across most settings, with REs consistently below $3\%$.
Similar to the observations under varying privacy-budget skewness, \textbf{Naive}, \textbf{Folded}, and \textbf{Folded-sum} provide essentially no utility across all settings, with REs consistently close to $100\%$, since the global bounds on which these baselines rely are overly conservative.
Turning to the oracle baselines, their utility deteriorates as $\lambda$ increases because their randomized thresholds are not adaptively selected.
In particular, when $\lambda = 2^7$, their REs exceed $24\%$ across all three privacy budget types, whereas our framework maintains REs below $1\%$.
These results demonstrate that our framework with adaptive threshold selection is substantially more robust to changes in the expected number of records per user than the baselines.
}

\subsection{Optimality of $\tau$}\label{app:optimal}
Since our method relies on selecting an optimal threshold $\tau$, we conduct an ablation study by manually varying $\tau$ over the range $2^{-9}$ to $2^{-2}$. 
We compare the resulting REs with those obtained by our framework on four datasets for \textit{Sum}, under the same privacy settings as in previous sections.

As shown in Table~\ref{tab:tau}, performance is highly sensitive to the choice of $\tau$ and can vary significantly across datasets: an inappropriate value just $2\times$ larger than the optimal can lead to more than a $500\times$ increase in RE. 
This underscores that selecting $\tau$ is critical for data utility, yet choosing an appropriate value remains challenging.
\bbd{From another perspective, these results also demonstrate that a single predefined global threshold cannot consistently balance truncation error and DP noise across different datasets.
}
Nevertheless, our framework consistently achieves REs within a small constant factor of those obtained with the best manually tuned $\tau$, as indicated by the optimality ratio, defined as $\text{Optimality Ratio} = \text{RE}(\text{Framework}) / \text{RE}(\text{manually set optimal } \tau) \times 100\%$.
This is expected, since our framework allocates a portion of the privacy budget to privately estimating $\tau$, whereas the baselines are strengthened by manually tuning $\tau$ to attain their best utility and are therefore not fully DP.
Overall, these results show that our framework can accurately estimate an appropriate $\tau$ for each case, thereby achieving high data utility.

\begin{table}[htbp]
  \caption{\bbd{Comparison of RE (\%) for \textit{Sum} achieved by our framework versus baselines with predefined $\tau$ values on four datasets.}}
  \label{tab:tau}
  \definecolor{mygray}{HTML}{E7E7E7}
  \setlength{\arrayrulewidth}{0.45pt}
  \setlength{\doublerulesep}{0.55pt}
  \renewcommand{\arraystretch}{0.95}
  \setlength{\tabcolsep}{6.0pt}

  \newcommand{\VC}[1]{\begin{tabular}[c]{@{}c@{}}#1\end{tabular}}
  \newcommand{\TopDbl}{\hhline{=====}}
  \newcommand{\MidDbl}{\hhline{=====}}

  \centering
  \footnotesize
  \makebox[\columnwidth][c]{%
  \resizebox{1\columnwidth}{!}{%
    \begin{tabular}{c|c|c|c|c}
      \TopDbl
      Dataset & CK Pay & Online Retail & Czech Bank & TPCH Order \\
      \MidDbl
      \VC{Framework}      & 0.1813 & 14.31 & 2.11  & 0.0378 \\
\hhline{-----}
$\tau=2^{-9}$      & 4.38  & 18.97  & 2.12  & 0.578   \\
\hhline{-----}
$\tau=2^{-8}$      & 2.01  & 14.17  & \cellcolor{mygray}1.114  & 0.1563   \\
\hhline{-----}
$\tau=2^{-7}$      & 1.494  & 11.94  & 1.780  & 0.1153   \\
\hhline{-----}
$\tau=2^{-6}$      & 0.726  & \cellcolor{mygray}10.86  & 14.84  & 0.0710   \\
\hhline{-----}
$\tau=2^{-5}$      & 0.486  & 12.55  & 39.6  & \cellcolor{mygray}0.0277   \\
\hhline{-----}
$\tau=2^{-4}$      & 0.1658  & 18.03  & 61.3  & 0.0355   \\
\hhline{-----}
$\tau=2^{-3}$      & \cellcolor{mygray}0.0849  & 26.9  & 77.3  & 0.800   \\
\hhline{-----}
$\tau=2^{-2}$      & 47.8  & 41.2  & 86.9  & 15.22   \\
\hhline{-----}
    Opt. Ratio (\%)  & 213.5   & 131.8  & 189.4  & 136.5   \\
      \TopDbl

    \end{tabular}%
  }%
}

\end{table}

\subsection{Privacy allocation}\label{app:allocation}

\begin{table}[htpb]
\caption{\bbd{Comparison of RE (\%) for our framework with different privacy-budget allocation ratios for count, sum, and max.}}\label{tab:alpha_results}

\definecolor{mygray}{HTML}{E7E7E7}
\setlength{\arrayrulewidth}{0.45pt}
\setlength{\doublerulesep}{0.55pt}
\renewcommand{\arraystretch}{1}
\setlength{\tabcolsep}{5.0pt}

\newcommand{\VC}[1]{\begin{tabular}[c]{@{}c@{}}#1\end{tabular}}
\newcommand{\TopDbl}{\hhline{==||====}}
\newcommand{\MidDbl}{\hhline{==||====}}

\centering
\arrayrulecolor{black}
\resizebox{\columnwidth}{!}{%
\begin{tabular}{c|c||c|c|c|c}
\TopDbl
\multicolumn{2}{c||}{Dataset} &
 CK Pay & NY Salary &
      Czech Bank & TPCH Lineitem \\
      \MidDbl

      \multirow{5}{*}{Count} &
      $\alpha=10\%$ &
      95.1 & 0.01169 &
      100.0 & 19.00 \\
      \hhline{~-||----}

      & $\alpha=30\%$ &
      0.00506 & 0.01191 &
      29.3 & 0.882 \\
      \hhline{~-||----}

      & $\alpha=50\%$ &
      \cellcolor{mygray}0.00302 & \cellcolor{mygray}0.01116 &
      16.35 & 0.881 \\
      \hhline{~-||----}

      & $\alpha=70\%$ &
      0.00714 & 0.01123 &
      12.67 & 0.882 \\
      \hhline{~-||----}

      & $\alpha=90\%$ &
      0.0236 & 0.01374 &
      \cellcolor{mygray}7.65 & \cellcolor{mygray}0.876 \\
      \MidDbl

      \multirow{5}{*}{Sum} &
      $\alpha=10\%$ &
      89.6 & \cellcolor{mygray}0.0497 &
      100.0 & 18.49 \\
      \hhline{~-||----}

      & $\alpha=30\%$ &
      \cellcolor{mygray}0.1519 & 0.0585 &
      14.83 & \cellcolor{mygray}0.856 \\
      \hhline{~-||----}

      & $\alpha=50\%$ &
      0.1813 & 0.1260 &
      \cellcolor{mygray}2.11 & 0.970 \\
      \hhline{~-||----}

      & $\alpha=70\%$ &
      0.292 & 0.1564 &
      3.06 & 0.874 \\
      \hhline{~-||----}

      & $\alpha=90\%$ &
      1.140 & 0.348 &
      11.36 & 1.442 \\
      \MidDbl

      \multirow{5}{*}{Max} &
      $\alpha=10\%$ &
      89.0 & \cellcolor{mygray}0.0692 &
      100.0 & \cellcolor{mygray}0.200 \\
      \hhline{~-||----}

      & $\alpha=30\%$ &
      \cellcolor{mygray}0.337 & 0.0877 &
      \cellcolor{mygray}0.728 & 0.422 \\
      \hhline{~-||----}

      & $\alpha=50\%$ &
      0.469 & 0.1258 &
      1.707 & 0.594 \\
      \hhline{~-||----}

      & $\alpha=70\%$ &
      0.772 & 0.207 &
      5.01 & 0.976 \\
      \hhline{~-||----}

      & $\alpha=90\%$ &
      2.33 & 0.621 &
      17.54 & 2.96 \\
      \TopDbl
    \end{tabular}%
}
\end{table}

\bbd{
We further investigate how privacy-budget allocation between the two stages of Algorithm~\ref{alg:Hybrid-PDP} affects utility. 
We introduce $\alpha\in(0\%,100\%)$ as the fraction of the privacy budget allocated to threshold selection, with the remaining $(100\%-\alpha)$ fraction allocated to the downstream query mechanism. Specifically, we modify Algorithm~\ref{alg:Hybrid-PDP} as follows. 
In Line~1, we use $\alpha\cdot\Phi_{\mathrm{rec}}$ and $\alpha\cdot\Phi_{\mathrm{user}}$ as the input privacy functions. 
Then, in Line~2, we rescale the selected threshold as $\tau_{i^\star} \gets \tau_{i^\star}^{\mathrm{half}}/\alpha$ to recover the real privacy threshold for the doubly-truncation. 
Finally, in Line~4, we use $\tau_{i^\star}\cdot(100\%-\alpha)/2$ as the uniform privacy budget for the downstream record-DP mechanism. 
By Theorem~\ref{thm:thld-select}, under this privacy-budget allocation, the threshold-selection stage satisfies $(\alpha\cdot\Phi_{\mathrm{rec}}, \alpha\cdot\Phi_{\mathrm{user}})$-Hybrid-PDP, while the downstream mechanism satisfies $((100\%-\alpha)\Phi_{\mathrm{rec}}, (100\%-\alpha)\Phi_{\mathrm{user}})$-Hybrid-PDP. Therefore, by sequential composition (Lemma~\ref{lem:hdp-comp}), the overall mechanism satisfies $(\Phi_{\mathrm{rec}}, \Phi_{\mathrm{user}})$-Hybrid-PDP.
Notably, when $\alpha=50\%$, this configuration reduces to the equal split adopted by default in our framework. We evaluate $\alpha \in \{10\%, 30\%, 50\%, 70\%, 90\%\}$ on four datasets with diverse privacy functions and across all three queries.
The observed preferences across query types are intuitive. 
The downstream mechanism for \textit{Count} is relatively simple, whereas those for \textit{Sum} and \textit{Max} are more complex and may therefore benefit more from allocating a larger fraction of the privacy budget to downstream query execution.
Consistent with this intuition, the results in Table~\ref{tab:alpha_results} show that \textit{Count} generally favors larger values of $\alpha$, while \textit{Sum} and \textit{Max} tend to favor smaller values.
Meanwhile, the best-performing allocation still varies across datasets, suggesting that no single value of $\alpha$ consistently performs the best.
Importantly, the equal allocation $\alpha=50\%$ performs consistently well across queries and datasets. 
In all 12 evaluated settings, its RE is within $3\times$ of that achieved by the best-performing allocation. Since selecting a dataset-specific optimal allocation ratio would require an additional tuning mechanism and incur extra privacy cost, these results support using $\alpha = 50\%$ as a default choice.
}

\section{Proofs of Theorems and Lemmas}\label{proof}

\subsection{Proof of Lemma~\ref{lem:privacy-sort}}

\begin{proof}
By symmetry of record-neighboring and of $d_{\mathrm{rec}}(\cdot,\cdot)$, it suffices to consider the case where
$D'$ is obtained from $D$ by deleting a single record $r^*$.
Let $u^*$ be the user owning $r^*$.

For every user $u \neq u^*$, the multi-set of records owned by $u$ is identical in $D$ and $D'$.
Hence the privacy-ordered list of records for $u$, and therefore the retained prefix after prefix-truncation, are identical in
$\overline D$ and $\overline D'$.
Thus only the records of $u^*$ can contribute to
$d_{\mathrm{rec}}(\overline D,\overline D')$.
Let
\[
r_1 \prec r_2 \prec \cdots \prec r_m
\]
be the records of $u^*$ in $D$, listed in privacy-based order.
Assume that ties in the privacy-based order are broken by a fixed deterministic
rule independent of the dataset.
By definition of this order,
\begin{equation}\label{eq:privacy-monotone}
\Phi_{\mathrm{rec}}(r_1)
\le
\Phi_{\mathrm{rec}}(r_2)
\le
\cdots
\le
\Phi_{\mathrm{rec}}(r_m).
\end{equation}
Assume that $r^* = r_t$ for some $t \in \{1,\ldots,m\}$, and define
\[
C_j := \sum_{i=1}^j \Phi_{\mathrm{rec}}(r_i),
\qquad
C_0 := 0.
\]
Let
\begin{equation}\label{eq:def-k-prefix}
k := \max\bigl\{j \in \{0,\ldots,m\} : C_j \le \Phi_{\mathrm{user}}(u^*)\bigr\}.
\end{equation}
Then the retained prefix of user $u^*$ in $\overline D$ is
\[
P := \{r_1,\ldots,r_k\}.
\]

If $k=m$, then all records of $u^*$ are retained in $\overline D$.
After deleting $r_t$, the retained prefix in $\overline D'$ is simply
$P \setminus \{r_t\}$.
Hence $\overline D$ and $\overline D'$ differ by exactly one record for user $u^*$, and therefore
\[
d_{\mathrm{rec}}(\overline D,\overline D') \le 1.
\]
So it remains to consider the case $k<m$.
By maximality of $k$,
\begin{equation}\label{eq:Sk}
C_k \le \Phi_{\mathrm{user}}(u^*) < C_{k+1}.
\end{equation}

For each $j \in \{0,\ldots,m\}$, define
\[
P'_j := \{r_1,\ldots,r_j\}\setminus\{r_t\}.
\]
Every prefix of the ordered records of $u^*$ in $D'$ can be written as $P'_j$ for some $j$.
Moreover,
\begin{equation}\label{eq:Pj-feasibility}
\sum_{r \in P'_j} \Phi_{\mathrm{rec}}(r)
=
\begin{cases}
C_j, & j<t,\\[4pt]
C_j - \Phi_{\mathrm{rec}}(r_t), & j\ge t.
\end{cases}
\end{equation}
Let $k'$ be the largest $j$ such that $P'_j$ is feasible, i.e.,
\[
\sum_{r \in P'_j} \Phi_{\mathrm{rec}}(r) \le \Phi_{\mathrm{user}}(u^*).
\]
Then the retained prefix of user $u^*$ in $\overline D'$ is
\[
P' := P'_{k'}.
\]

We now show that $k' \in \{k,k+1\}$.

First, $P'_k$ is always feasible.
If $t>k$, then $P'_k = P$, and its total privacy cost is $C_k \le \Phi_{\mathrm{user}}(u^*)$.
If $t\le k$, then
\[
\sum_{r \in P'_k} \Phi_{\mathrm{rec}}(r)
=
C_k - \Phi_{\mathrm{rec}}(r_t)
\le
C_k
\le
\Phi_{\mathrm{user}}(u^*).
\]
Therefore,
\begin{equation}\label{eq:jstar-lb}
k' \ge k.
\end{equation}

Second, no $P'_j$ with $j\ge k+2$ can be feasible.
Indeed, if $j<t$, then $j>k$, so by (\ref{eq:Sk}),
\[
\sum_{r \in P'_j} \Phi_{\mathrm{rec}}(r)
=
C_j
\ge
C_{k+1}
>
\Phi_{\mathrm{user}}(u^*).
\]
If $j\ge t$, then by (\ref{eq:privacy-monotone}),
\[
\Phi_{\mathrm{rec}}(r_t) \le \Phi_{\mathrm{rec}}(r_j),
\]
and hence
\[
\begin{aligned}
\sum_{r \in P'_j} \Phi_{\mathrm{rec}}(r)
&=
C_j - \Phi_{\mathrm{rec}}(r_t) \ge
C_j - \Phi_{\mathrm{rec}}(r_j) \\
&=
C_{j-1} \ge
C_{k+1} >
\Phi_{\mathrm{user}}(u^*).
\end{aligned}
\]
Thus,
\begin{equation}\label{eq:jstar-ub}
k' \le k+1.
\end{equation}

Combining (\ref{eq:jstar-lb}) and (\ref{eq:jstar-ub}), we obtain
\begin{equation*}
    k' \in \{k,k+1\}.
\end{equation*}

Therefore, compared with
\[
P = \{r_1,\ldots,r_k\},
\]
the retained prefix $P'$ is obtained by possibly removing the deleted record $r_t$, and possibly adding the single next record $r_{k+1}$.
Equivalently, one can transform $P$ into $P'$ using at most one record deletion and at most one record insertion.

Since all users other than $u^*$ have identical retained prefixes in $\overline D$ and $\overline D'$, it follows that
\[
d_{\mathrm{rec}}(\overline D,\overline D') \le 2.
\]
This completes the proof.
\end{proof}

\subsection{Proof of Lemma~\ref{lem:domin}}
\begin{proof}
Fix any user $u$. We consider two cases.

First, suppose that $u \in D$. By the definition of prefix-truncation, for user $u$ we retain the longest prefix of $D(u)$, under the privacy-based ordering, whose cumulative record-level privacy budget does not exceed $\Phi_{\mathrm{user}}(u)$. Therefore, the retained records satisfy
\[
\sum_{r \in \overline D(u)} \Phi_{\mathrm{rec}}(r)
\leq
\Phi_{\mathrm{user}}(u),
\qquad \forall u \in D.
\]

Otherwise, suppose that $u \notin D$. Then $D(u)=\emptyset$, and hence prefix-truncation also gives $\overline D(u)=\emptyset$. Thus,
\[
\sum_{r \in \overline D(u)} \Phi_{\mathrm{rec}}(r)
=
0
<
\Phi_{\mathrm{user}}(u),
\qquad \forall u \notin D.
\]

Combining the two cases, for every user $u \in \mathcal U$,
\[
\sum_{r \in \overline D(u)} \Phi_{\mathrm{rec}}(r)
\leq
\Phi_{\mathrm{user}}(u).
\]
This completes the proof.
\end{proof}

\subsection{Proof of Lemma~\ref{lem:privacy-sort-dt}}

\begin{proof}
Since $D\sim_r D'$, without loss of generality, suppose that $D'$ is obtained from $D$ by adding one record $r^*$ for some user $u^*$. The deletion case is symmetric.

Let $D_{\tau}$ and $D'_{\tau}$ be the datasets obtained from $D$ and $D'$ after the first truncation step. We consider two cases.

First, suppose that $\Phi_{\mathrm{rec}}(r^*)<\tau$. Then $r^*$ is removed in the first truncation step. Hence,
\[
D_{\tau}=D'_{\tau}.
\]

Otherwise, suppose that $\Phi_{\mathrm{rec}}(r^*)\geq \tau$. Then $r^*$ is retained. Since no other record is changed by the first truncation step, $D_{\tau}$ and $D'_{\tau}$ still differ by exactly one record. Thus,
\[
d_{\mathrm{rec}}(D_{\tau},D'_{\tau})\leq 1.
\]

The privacy upgrade step only changes the record-level privacy budget of each retained record to $\tau$ and does not change the underlying record set. Therefore, after the first truncation and privacy upgrade, the two datasets remain identical or record-neighboring.

It remains to consider the final prefix-truncation step. After the privacy upgrade, every retained record has the same record-level privacy budget $\tau$. Hence, sorting records by nondecreasing record-level privacy budget is equivalent to using the fixed natural order as the tie-breaking order.

For each user $u$, the number of retained records is
\[
\kappa(u)
=
\min\left(
\left\lfloor \frac{\Phi_{\mathrm{user}}(u)}{\tau}\right\rfloor,
|D_{\tau}(u)|
\right),
\]
which is exactly the length of the longest prefix whose cumulative upgraded record-level privacy budget is at most $\Phi_{\mathrm{user}}(u)$.

Therefore, the second stage of doubly-truncation is precisely privacy-based ordering prefix-truncation applied to the datasets obtained after the first truncation and privacy upgrade. By Lemma~\ref{lem:privacy-sort}, applying such prefix-truncation to two record-neighboring datasets can increase their record-level distance to at most $2$. Consequently,
\[
d_{\mathrm{rec}}(\overline{\overline D},\overline{\overline {D'}})\leq 2.
\]
This completes the proof.
\end{proof}

\subsection{Proof of Lemma~\ref{lem:domin-dt}}

\begin{proof}
Fix any user $u$. We consider two cases.

First, suppose that $u \in D$. By the definition of doubly-truncation, we first retain only the records
\[
D_{\tau}(u)
=
\{r\in D(u): \Phi_{\mathrm{rec}}(r)\geq \tau\}.
\]
For every record $r\in D_{\tau}(u)$, its privacy budget after doubly-truncation is upgraded to $\tau$.
Then, doubly-truncation retains the first $\kappa(u)$ records of $D_{\tau}(u)$, where
\[
\kappa(u)
=
\min\left(
\left\lfloor \frac{\Phi_{\mathrm{user}}(u)}{\tau}\right\rfloor,
|D_{\tau}(u)|
\right).
\]
Therefore, since every retained record has upgraded privacy budget $\tau$, we have
\[
\sum_{r\in \overline{\overline D}(u)} \tau
=
|\overline{\overline D}(u)|\cdot \tau
=
\kappa(u)\tau\leq\Phi_{\mathrm{user}}(u),\qquad \forall u\in D.
\]

Otherwise, suppose that $u\notin D$. Then $D(u)=\emptyset$, and hence $D_{\tau}(u)=\emptyset$. Therefore, doubly-truncation also gives
$\overline{\overline D}(u)=\emptyset$. Thus,
\[
\sum_{r\in \overline{\overline D}(u)} \tau
=
0
\leq
\Phi_{\mathrm{user}}(u),
\qquad \forall u\notin D.
\]

Combining the two cases, for every user $u\in \mathcal U$,
\[
\sum_{r\in \overline{\overline D}(u)} \tau
\leq
\Phi_{\mathrm{user}}(u).
\]
This completes the proof.
\end{proof}

\subsection{Proof of Lemma~\ref{lem:score}}

\begin{proof}
By symmetry, it suffices to consider the case where $D'$ is obtained from $D$ by deleting one record $r^*$.

Fix $\tau_i$. In the first truncation step, records with $\Phi_{\mathrm{rec}}(r) < \tau_i$ are removed independently record by record. Hence, after this step, the resulting datasets are either identical or still record-neighboring.

If they are identical, then applying the second step yields
\[
\overline{\overline D}_i=\overline{\overline {D'}}_i,
\]
and thus
\[
|S_i(D)-S_i(D')|=0.
\]

Otherwise, the two intermediate datasets are record-neighboring. Applying prefix-truncation in the second step, the argument in the proof of Lemma~\ref{lem:privacy-sort} implies that deleting one record can remove that record from the retained prefix and admit at most one additional record. Therefore,
\[
\bigl||\overline{\overline D}_i|-|\overline{\overline {D'}}_i|\bigr|\le 1.
\]
Since
\[
S_i(D)=|\overline{\overline D}_i|- \frac{3\ell}{\tau_i} \ln(\ell/\beta),
\qquad
S_i(D')=|\overline{\overline {D'}}_i|- \frac{3\ell}{\tau_i} \ln(\ell/\beta),
\]
we obtain
\[
|S_i(D)-S_i(D')|\le 1.
\]
This completes the proof.
\end{proof}

\subsection{Proof of Theorem~\ref{thm:thld-select}}

\begin{proof}\label{prf:th-sl}
Recall that Algorithm~\ref{alg:thld-select} sets
\begin{equation*}
\ell = \left\lceil \log\!\left( \hat{\varepsilon}/\check{\varepsilon} \right) \right\rceil+1,
\quad
\tau_i = 2^{i-1}\check{\varepsilon},
\end{equation*}
and releases
\begin{equation*}
\widetilde{S}_i
=
\bigl|\overline{\overline D}_i\bigr|
-\frac{3\ell}{\tau_i}\ln\!\frac{\ell}{\beta}
+\mathrm{Lap}\!\left(\frac{\ell}{\tau_i}\right).
\end{equation*}

\paragraph{Privacy.}
We prove that Algorithm~\ref{alg:thld-select} is
$(\Phi_{\mathrm{rec}},\,\Phi_{\mathrm{user}})$-Hybrid-PDP.
Recall that, for each iteration $i$, $S_i$ is the non-noisy score computed after
doubly-truncation with threshold $\tau_i$, and the released privatized score is
\begin{equation*}
\widetilde S_i
=
\bigl|\overline{\overline D}_i\bigr|
-\frac{3\ell}{\tau_i}\ln\!\frac{\ell}{\beta}
+
\mathrm{Lap}\!\left(\frac{\ell}{\tau_i}\right).
\end{equation*}
The deterministic bias term does not affect privacy.

\noindent\textbf{Record-level privacy guarantee for each iteration.} We first consider the record-level guarantee for a fixed iteration $i$.
Fix any record $r$. If $\tau_i>\Phi_{\mathrm{rec}}(r)$, then by the
record-level truncation in doubly-truncation, record $r$ is removed before
forming the truncated dataset. Hence the distribution of $\widetilde S_i$
is identical whether or not $r$ is present, and the privacy loss for $r$
in this iteration is \textbf{zero}.

Otherwise, if $\tau_i \le \Phi_{\mathrm{rec}}(r)$, then by
Lemma~\ref{lem:score}, the record-level sensitivity of $S_i$ is at most one,
i.e., $\mathrm{GS}_{S_i} = 1$.

Therefore, by the Laplace mechanism, adding
$\mathrm{Lap}(\ell/\tau_i)$ ensures that releasing $\widetilde S_i$
satisfies $(\tau_i/\ell)$-record-level DP with respect to record $r$. Since
$\tau_i\le \Phi_{\mathrm{rec}}(r)$, we have
$\tau_i/\ell\le\Phi_{\mathrm{rec}}(r)/\ell$.
Thus, the $i$-th iteration satisfies
$(\Phi_{\mathrm{rec}}/\ell)$-record-level privacy guarantee.

\noindent\textbf{User-level privacy guarantee for each iteration.} We next verify the user-level guarantee for the same fixed iteration $i$.
Fix any user $u$, and consider two user-neighboring datasets that differ only
in the records contributed by user $u$. Recall that $\overline{\overline D}_i(u)$ is the retained records of user
$u$ after doubly-truncation with threshold
$\tau_i$. Since the score $S_i$ is the size of the truncated dataset, the
change in $S_i$ caused by adding or removing user $u$ is at most
$|\overline{\overline D}_i(u)|$. Therefore, releasing $\widetilde S_i$ with
Laplace noise scale $\ell/\tau_i$ incurs user-level privacy loss at most
\begin{equation*}
\frac{|\overline{\overline D}_i(u)|}{\ell/\tau_i}
=
\frac{1}{\ell}
\sum_{r\in \overline{\overline D}_i(u)} \tau_i .
\end{equation*}
By Lemma~\ref{lem:domin-dt}, doubly-truncation guarantees that
$\sum_{r\in \overline{\overline D}_i(u)} \tau_i\le\Phi_{\mathrm{user}}(u)$.
Hence the privacy loss for user $u$ in the $i$-th iteration is at most $\Phi_{\mathrm{user}}(u)/\ell$.
Thus, the $i$-th iteration satisfies the
$(\Phi_{\mathrm{user}}/\ell)$-user-level privacy guarantee. 

\noindent\textbf{Composition across iterations.}
Combining this with
the record-level argument above, each iteration is
$\left(\Phi_{\mathrm{rec}}/\ell,\Phi_{\mathrm{user}}/\ell\right)\text{-Hybrid-PDP}$.
Applying the sequential composition (Lemma~\ref{lem:hdp-comp}) over the $\ell$ iterations, the release
of the full scores
$(\widetilde S_1,\ldots,\widetilde S_\ell)$
satisfies
$(\Phi_{\mathrm{rec}},\Phi_{\mathrm{user}})$
-Hybrid-PDP.
Finally, the selected index
$i^\star=\arg\max_{1\le i\le \ell}\widetilde S_i$ and the output threshold
$\tau_{i^\star}$ are post-processing of the released scores. Therefore,
by Lemma~\ref{lem:hdp-post}, releasing $\tau_{i^\star}$ also satisfies
$(\Phi_{\mathrm{rec}},\Phi_{\mathrm{user}})$-Hybrid-PDP.

\paragraph{Utility.}

By the Laplace tail bound in Lemma~\ref{lem:laplace}, for each iteration $i$, with probability $1-\beta/\ell$,
\begin{equation}\label{eq:good-event}
-\frac{\ell}{\tau_i}\ln\!\frac{\ell}{\beta}
\;\le\;
\mathrm{Lap}\!\left(\frac{\ell}{\tau_i}\right)
\;\le\;
\frac{\ell}{\tau_i}\ln\!\frac{\ell}{\beta}.
\end{equation}
By applying the union bound over iterations $i=1,2,\ldots,\ell$, with probability at least $1-\beta$, the following holds:
\begin{equation}\label{eq:good-event}
-\frac{\ell}{\tau_i}\ln\!\frac{\ell}{\beta}
\;\le\;
\mathrm{Lap}\!\left(\frac{\ell}{\tau_i}\right)
\;\le\;
\frac{\ell}{\tau_i}\ln\!\frac{\ell}{\beta},
\quad i=1,2,\ldots,\ell .
\end{equation}

With \eqref{eq:good-event}, we prove claims (i) and (ii) separately.

\noindent Recall (i) $\tau_{i^\star} \ge \tfrac{1}{2}\,\varepsilon_{\min}(D)$:

Let $k$ be the index such that
\begin{equation*}
\varepsilon_{\min}(D) \in [\tau_k,\,2\tau_k).
\end{equation*}
By the definition of $\varepsilon_{\min}(D)$, for every $j\le k$, no record is removed
by doubly-truncation. Hence
\begin{equation*}
\bigl|\overline{\overline D}_j\bigr| = |D|
\quad\forall j\le k.
\end{equation*}

Fix any $j\le k-1$. Using the above and expanding
$\widetilde S_k-\widetilde S_j$, we have
\begin{equation*}
\begin{aligned}
\widetilde S_k-\widetilde S_j
&=
\bigl|\overline{\overline D}_k\bigr|
-
\bigl|\overline{\overline D}_j\bigr|
-\left(\frac{3\ell}{\tau_k}-\frac{3\ell}{\tau_j}\right)
\ln\!\frac{\ell}{\beta}
+
\mathrm{Lap}\!\left(\frac{\ell}{\tau_k}\right)
-
\mathrm{Lap}\!\left(\frac{\ell}{\tau_j}\right)
\\
&=
-\left(\frac{3\ell}{\tau_k}-\frac{3\ell}{\tau_j}\right)
\ln\!\frac{\ell}{\beta}
+
\mathrm{Lap}\!\left(\frac{\ell}{\tau_k}\right)
-
\mathrm{Lap}\!\left(\frac{\ell}{\tau_j}\right)
\\
&\ge
-\left(\frac{3\ell}{\tau_k}-\frac{3\ell}{\tau_j}\right)
\ln\!\frac{\ell}{\beta}
-
\left(\frac{\ell}{\tau_k}+\frac{\ell}{\tau_j}\right)
\ln\!\frac{\ell}{\beta}
\\
&=
\left(\frac{2\ell}{\tau_j}-\frac{4\ell}{\tau_k}\right)
\ln\!\frac{\ell}{\beta}.
\end{aligned}
\end{equation*}
Since $j\le k-1$ implies $\tau_j\le \tau_k/2$, we have
$1/\tau_j\ge 2/\tau_k$, and thus
\begin{equation*}
\left(\frac{2\ell}{\tau_j}-\frac{4\ell}{\tau_k}\right)
\ln\!\frac{\ell}{\beta}
\ge 0.
\end{equation*}
Therefore, $\widetilde S_k\ge \widetilde S_j$ for all $j\le k-1$.
That is, no index $j<k$ has a larger privatized score than index $k$.
Therefore,
\begin{equation*}
    \tau_{i^\star}
    \ge
    \tau_k
    >
    \tfrac{1}{2}\,\varepsilon_{\min}(D),
\end{equation*}
which proves claim~(i).

\noindent Recall (ii) $\bigl|\,|D| - |\overline{\overline D}_{i^\star}|\bigr| = O\!\left(\dfrac{1}{\varepsilon_{\min}(D)}\log(\hat{\varepsilon}/\check{\varepsilon}) \log\!\frac{\log(\hat{\varepsilon}/\check{\varepsilon})}{\beta}\right)$, 
where $\overline{\overline D}_{i^\star}$ is obtained by applying doubly-truncation to $D$ with threshold $\tau_{i^\star}$: 

Since $i^\star$ is the index with the best score, $\widetilde S_{i^\star}$, we have
\begin{equation*}
\widetilde S_{i^\star} \ge \widetilde S_k.
\end{equation*}
Expanding both $\widetilde S_{i^\star}$ and $\widetilde S_k$, and using $\bigl|\overline{\overline D}_k\bigr|=|D|$, we obtain
\begin{equation*}
\begin{aligned}
\bigl|\overline{\overline D}_{i^\star}\bigr|
-\frac{3\ell}{\tau_{i^\star}}\ln\!\frac{\ell}{\beta}
+\mathrm{Lap}\!\left(\frac{\ell}{\tau_{i^\star}}\right)
&\ge
|D|
-\frac{3\ell}{\tau_k}\ln\!\frac{\ell}{\beta}
+\mathrm{Lap}\!\left(\frac{\ell}{\tau_k}\right).
\end{aligned}
\end{equation*}
Rearranging yields
\begin{equation*}
\begin{aligned}
|D|-\bigl|\overline{\overline D}_{i^\star}\bigr|
&\le
\mathrm{Lap}\!\left(\frac{\ell}{\tau_{i^\star}}\right)
-
\mathrm{Lap}\!\left(\frac{\ell}{\tau_k}\right)
+
\left(\frac{3\ell}{\tau_k}
-
\frac{3\ell}{\tau_{i^\star}}\right)
\ln\!\frac{\ell}{\beta}.
\end{aligned}
\end{equation*}
Applying \eqref{eq:good-event} to both Laplace terms gives
\begin{equation*}
\begin{aligned}
|D|-\bigl|\overline{\overline D}_{i^\star}\bigr|
&\le
\left(\frac{\ell}{\tau_{i^\star}}+\frac{\ell}{\tau_k}\right)
\ln\!\frac{\ell}{\beta}
+
\left(\frac{3\ell}{\tau_k}
-
\frac{3\ell}{\tau_{i^\star}}\right)
\ln\!\frac{\ell}{\beta}
\\
&=
\left(\frac{4\ell}{\tau_k}
-
\frac{2\ell}{\tau_{i^\star}}\right)
\ln\!\frac{\ell}{\beta}
\\
&\le
\frac{4\ell}{\tau_k}\ln\!\frac{\ell}{\beta}.
\end{aligned}
\end{equation*}
Finally, since $\varepsilon_{\min}(D)\in[\tau_k,2\tau_k)$, we have
$1/\tau_k\le 2/\varepsilon_{\min}(D)$, and thus
\begin{equation*}
\begin{aligned}
|D|-\bigl|\overline{\overline D}_{i^\star}\bigr|
&\le
\frac{8\ell}{\varepsilon_{\min}(D)}
\ln\!\frac{\ell}{\beta}.
\end{aligned}
\end{equation*}
Noting that
$\ell=\left\lceil\log(\hat{\varepsilon}/\check{\varepsilon})\right\rceil+1$,
this implies
\begin{equation*}
|D|-\bigl|\overline{\overline D}_{i^\star}\bigr|
=
O\!\left(
\frac{\ell}{\varepsilon_{\min}(D)}
\log\!\frac{\ell}{\beta}
\right)
=
O\!\left(
\frac{1}{\varepsilon_{\min}(D)}\log(\hat{\varepsilon}/\check{\varepsilon})
\log\!\frac{\log(\hat{\varepsilon}/\check{\varepsilon})}{\beta}
\right),
\end{equation*}
which proves claim~(ii).
\end{proof}

\subsection{Proof of Theorem~\ref{thm:utility}}
\begin{proof}
We prove the theorem in two parts: privacy and utility.

\paragraph{Privacy.}
We first prove that Algorithm~\ref{alg:Hybrid-PDP} satisfies
$(\Phi_{\mathrm{rec}},\Phi_{\mathrm{user}})$-Hybrid-PDP.

\noindent(1) In Stage~1, the algorithm invokes
\[
\textsc{ThresholdSelection}
(D,\Phi_{\mathrm{rec}}/2,\Phi_{\mathrm{user}}/2,\beta/2).
\]
By Theorem~\ref{thm:thld-select}, this step satisfies
$(\Phi_{\mathrm{rec}}/2,\Phi_{\mathrm{user}}/2)$-Hybrid-PDP.
The subsequent assignment
\begin{equation*}
    \tau_{i^\star} = 2\tau_{i^\star}^{\mathrm{half}}
\end{equation*}
is deterministic post-processing of the selected threshold and
therefore does not consume any additional privacy budget.

\noindent(2) In Stage~2, the algorithm computes
\[
\overline{\overline D}_{i^\star}
\gets
\textsc{Doubly-Truncation}
(D,\Phi_{\mathrm{rec}},\Phi_{\mathrm{user}},\tau_{i^\star})
\]
and then invokes
\[
\mathcal M
\bigl(
\overline{\overline D}_{i^\star},
\tau_{i^\star}/4,
\beta/2
\bigr).
\]
\noindent\textbf{Record-level guarantee in Stage~2.} For any record $r$, if $\Phi_{\mathrm{rec}}(r)<\tau_{i^\star}$,
then $r$ is discarded by doubly-truncation and consumes no privacy
budget in Stage~2. Otherwise, $\Phi_{\mathrm{rec}}(r)\ge\tau_{i^\star}$.
By Lemma~\ref{lem:privacy-sort-dt}, after doubly-truncation, the distance between two record-level neighboring datasets is at most 2.
Since $\mathcal M$ is invoked with privacy budget $\tau_{i^\star}/4$, by group privacy (Lemma~\ref{lem:group}), the privacy consumption of any record $r$ in Stage~2 is at most
\[
2\cdot \frac{\tau_{i^\star}}{4}
=
\frac{\tau_{i^\star}}{2}
\le
\frac{\Phi_{\mathrm{rec}}(r)}{2}.
\]
Hence Stage~2 satisfies the $\Phi_{\mathrm{rec}}/2$ record-level
privacy requirement.

\noindent\textbf{User-level guarantee for Stage~2.}
By Lemma~\ref{lem:domin-dt}, after doubly-truncation with threshold
$\tau_{i^\star}$, the retained records of any user $u$ have total
upgraded record-level budget at most $\Phi_{\mathrm{user}}(u)$.
Since each retained record is assigned privacy budget $\tau_{i^\star}$,
after doubly-truncation, two user-level neighboring datasets that differ in user $u$
can differ in at most
\begin{equation*}
    2 \cdot \frac{\Phi_{\mathrm{user}}(u)}{\tau_{i^\star}}
\end{equation*}
retained records. By group privacy (Lemma~\ref{lem:group}), invoking $\mathcal{M}$ with budget
$\tau_{i^\star}/4$ therefore incurs user-level privacy loss at most
\begin{equation*}
 2 \cdot 
    \frac{\Phi_{\mathrm{user}}(u)}{\tau_{i^\star}}
    \cdot
    \frac{\tau_{i^\star}}{4}
    =
    \frac{\Phi_{\mathrm{user}}(u)}{2}.
\end{equation*}
Thus Stage~2 satisfies the $\Phi_{\mathrm{user}}/2$ user-level
privacy requirement.

Therefore, Stage~2 is
$(\Phi_{\mathrm{rec}}/2,\Phi_{\mathrm{user}}/2)$-Hybrid-PDP.

By the sequential composition of Hybrid-PDP (Lemma~\ref{lem:hdp-comp}), the whole
algorithm satisfies
$(\Phi_{\mathrm{rec}},\Phi_{\mathrm{user}})$-Hybrid-PDP.

\paragraph{Utility.}
Let $D'$ denote the selected retained dataset
$\overline{\overline{D}}_{i^\star}$ in Algorithm~\ref{alg:Hybrid-PDP}.
Since the threshold selection in our framework runs 
with privacy functions $\Phi_{\mathrm{rec}}/2$ and $\Phi_{\mathrm{user}}/2$ and failure rate $\beta/2$, 
Theorem~\ref{thm:thld-select} implies that, with probability at least
$1-\beta/2$, the selected threshold
$\tau_{i^\star}^{\mathrm{half}}$ satisfies
\[
\tau_{i^\star}^{\mathrm{half}}
\ge
\frac{1}{2}\cdot \frac{\varepsilon_{\min}(D)}{2}
=
\frac{\varepsilon_{\min}(D)}{4}.
\]
Since Algorithm~\ref{alg:Hybrid-PDP} sets
$\tau_{i^\star}=2\tau_{i^\star}^{\mathrm{half}}$, we have
\[
\tau_{i^\star}\ge \frac{1}{2}\varepsilon_{\min}(D).
\]
Moreover, the same theorem gives
\begin{equation}\label{eq:thld-good}
\bigl|\,|D|-|D'|\,\bigr|
=
O\!\left(
\frac{1}{\varepsilon_{\min}(D)}\log(\hat{\varepsilon}/\check{\varepsilon})
\log\!\frac{\log(\hat{\varepsilon}/\check{\varepsilon})}{\beta}
\right).
\end{equation}

With $D'$, Algorithm~\ref{alg:Hybrid-PDP} runs
$\mathcal{M}$ with privacy budget $\tau_{i^\star}/4$ and failure
probability $\beta/2$ to output $\widetilde{Q}(D)$. Since
\[
\frac{\tau_{i^\star}}{4}
\ge
\frac{\varepsilon_{\min}(D)}{8},
\]
by the definition of $\mathrm{Err}_{\mathcal{M}}(\cdot)$, with
probability at least $1-\beta/2$,
\begin{equation}\label{eq:M-good}
\bigl|\widetilde{Q}(D)-Q(D')\bigr|
\le
\mathrm{Err}_{\mathcal{M}}
\!\left(D',\,\varepsilon_{\min}(D)/8,\,\beta/2\right).
\end{equation}

By a union bound over
\eqref{eq:thld-good} and \eqref{eq:M-good}, both hold simultaneously
with probability at least $1-\beta$. By the triangle
inequality,
\begin{equation*}
\begin{aligned}
\bigl|\widetilde{Q}(D)-Q(D)\bigr|
&\le
\bigl|Q(D)-Q(D')\bigr|
+
\bigl|\widetilde{Q}(D)-Q(D')\bigr| \\
&\le
\bigl|Q(D)-Q(D')\bigr|
+
\mathrm{Err}_{\mathcal{M}}
\!\left(D',\,\varepsilon_{\min}(D)/8,\,\beta/2\right),
\end{aligned}
\end{equation*}
where $D'$ is obtained from $D$ by removing
\[
O\!\left(
\frac{1}{\varepsilon_{\min}(D)}\log(\hat{\varepsilon}/\check{\varepsilon})
\log\!\frac{\log(\hat{\varepsilon}/\check{\varepsilon})}{\beta}
\right)
\]
records.
\end{proof}

\section{Addressing Privacy Leakage from Public Privacy Budgets Using Multiple Copies}
\label{app:private-budgets}

\gbd{In both standard PDP and our Hybrid-PDP model, the privacy budget assigned to each user or record is public and may be correlated with sensitive information.
For example, employees with lower performance scores may prefer smaller privacy budgets, as revealing their scores could have greater negative effects on them.
Consequently, directly publishing an employee's actual privacy budget may indirectly reveal their score.
To prevent such disclosures, we construct multiple public copies for each potential user or record, assigning each copy a different budget from a predefined global range.
We publish each copy's budget without revealing which copy corresponds to the individual's actual privacy budget.
These copies are constructed independently of the private dataset.
In addition, these copies cover all possible privacy budgets and thus do not restrict the domain of privacy budgets.
For example, suppose an individual $A$ (either a user or a record) has an actual privacy budget of $0.100$, and the allowable privacy budgets are $\{0.001, 0.002, 0.003, \ldots, 1.000\}$\footnote{\gbd{In practice, the budget granularity can be chosen to be sufficiently
fine to cover all privacy budget values supported by the system.}}.
We publish $1000$ copies, $A_1,\ldots,A_{1000}$, with $\Phi(A_i) = 0.001i$.
Only $A_{100}$ is included in the private dataset $D$.
While the trusted curator knows which copy is selected, external analysts can observe all public copies without knowing which one corresponds to $A$'s actual privacy budget.
The trusted curator executes the Hybrid-PDP mechanism, which internally incorporates the downstream record-DP mechanism, and releases only the privatized query outputs.
Table~\ref{tab:budget-visibility} summarizes the information observable to external analysts under this design.}

\begin{table}[ht]
    \centering
    \caption{\gbd{Information directly observable to external analysts.
    Here, $A$ denotes an actual user or record, and $A_i$ denotes
    its $i$-th public copy.}}
    \label{tab:budget-visibility}
    \begin{tabular}{lc}
        \toprule
        \gbd{Information} & \gbd{Observable} \\
        \midrule
        \gbd{Each public copy's budget $\Phi(A_i)$} & \gbd{Yes} \\
        \gbd{Actual privacy budget of $A$} & \gbd{No} \\
        \gbd{Presence of $A$ in $D$} & \gbd{No} \\
        \gbd{Sensitive values associated with $A$} & \gbd{No} \\
        \gbd{Released query results $\mathcal{M}(D)$} & \gbd{Yes} \\
        \bottomrule
    \end{tabular}
\end{table}

\gbd{For each public copy $A_i$, Hybrid-PDP protects its presence or absence in $D$ under the assigned privacy budget $\Phi(A_i)$.
Thus, external analysts cannot determine from the privatized query outputs whether any particular copy is included in $D$.
Moreover, since the public copies and their associated budgets are constructed independently of the individual's actual privacy budget, they reveal no information about it.}